\documentclass[prd,twocolumn,showpacs,preprintnumbers,amsmath,amssymb]{revtex4}
\usepackage{graphicx}% Include figure files
\usepackage{dcolumn}% Align table columns on decimal point
\usepackage{bm}% bold math
\usepackage{wrapfig} 
\usepackage{color} 
\newcommand{\lsim}{\raisebox{-0.13cm}{~\shortstack{$<$ \\[-0.07cm] $\sim$}}~} 
\newcommand{\gsim}{\raisebox{-0.13cm}{~\shortstack{$>$ \\[-0.07cm] $\sim$}}~} 
 
\newcommand{\lra}{\longrightarrow} 
\newcommand{\tb}{\tan\beta} 
\newcommand{\beq}{\begin{eqnarray}} 
\newcommand{\eeq}{\end{eqnarray}}

\begin{document}

%\preprint{APS/123-QED}
\preprint{CERN--PH--TH/2013--053}\vspace*{0.5cm}

\title{Supersymmetric Heavy Higgs Bosons at the LHC}

\author{Alexandre Arbey}
\email{alexandre.arbey@ens-lyon.fr}
\affiliation{Centre de Recherche Astrophysique de Lyon, Observatoire de Lyon,\\
\mbox{Saint-Genis Laval Cedex, F-69561, France; CNRS, UMR 5574;}\\ 
Ecole Normale Sup\'erieure de Lyon, France;\\
\mbox{Universit\'e de Lyon, Universit\'e Lyon 1, F-69622~Villeurbanne Cedex, France}\\
\mbox{and CERN Theory Division, CH-1211 Geneva, Switzerland}}%

\author{Marco Battaglia}
\email{MBattaglia@lbl.gov}
\affiliation{University of California at Santa Cruz, \\
Santa Cruz Institute of Particle Physics, CA 94720, USA\\
and CERN, CH-1211 Geneva, Switzerland}%

\author{Farvah Mahmoudi}%
\email{mahmoudi@in2p3.fr}
\affiliation{Clermont Universit\'e, Universit\'e Blaise Pascal, CNRS/IN2P3,\\
LPC, BP 10448, F-63000 Clermont-Ferrand, France\\
and CERN Theory Division, CH-1211 Geneva, Switzerland}%

% \date{\today}% It is always \today, today,
             %  but any date may be explicitly specified

\begin{abstract}
\vspace*{0.5cm}The search for heavy Higgs bosons is an essential step in the exploration of the 
Higgs sector and in probing the Supersymmetric parameter space. This paper discusses the constraints 
on the $M_A$ and $\tan \beta$ parameters derived from the bounds on the different decay channels of the 
neutral $H$ and $A$ bosons accessible at the LHC, in the framework of the phenomenological MSSM. 
The implications from the present LHC results and the expected sensitivity of the 14~TeV data are 
discussed in terms of the coverage of the $[M_A - \tan \beta]$ plane. New channels becoming important 
at 13 and 14~TeV for low values of $\tan \beta$ are characterised in terms of their kinematics and the 
reconstruction strategies. The effect of QCD systematics, SUSY loop effects and decays into pairs of 
SUSY particles on these constraints are discussed in details. 
\end{abstract}

% PACS, the Physics and Astronomy
                        % Classification Scheme.
%\keywords{Suggested keywords}%Use showkeys class option if keyword
                              %display desired
\maketitle

\section{Introduction}

With the observation of a light Higgs-like particle by the ATLAS and CMS experiments at the 
LHC~\cite{Chatrchyan:2012ufa,Aad:2012tfa}, 
the detailed exploration of the Higgs sector becomes one of the most compelling programs of 
collider physics. In particular, understanding whether this sector extends beyond that of 
the Standard Model (SM) and heavier Higgs bosons exist is of crucial importance for the 
viability of several models of new physics beyond the Standard  Model, {\it in primis} of 
Supersymmetry (SUSY). This question can be answered either through a precision study of the 
couplings of the lightest boson, $h$, or by direct searches of the additional, heavier states 
which characterise extended Higgs models. 

The LHC experiments have not only observed a light 
state and obtained the first determination of its decay rates to $\gamma \gamma$, $WW$ and $ZZ$. 
They have also performed several searches directly probing the possible production of heavy Higgs 
bosons and other searches, which can now be re-interpreted in order to set constraints on the 
production and decays of neutral heavy Higgs states. 
However, these data are still largely fragmentary. 

Several studies of the MSSM heavy Higgs sector at LHC results have already been 
performed~\cite{Baglio:2011xz,Carena:2011fc,Christensen:2012ei,Chang:2012zf,Carena:2013qia}.
This paper intends to provide a comprehensive assessment of the present status and the future 
perspectives for the constraints on the MSSM Higgs sector parameters, from the identification 
of the  main processes relevant to the LHC searches to a systematic 
study of the exclusion limits derived from the combination of the LHC results, in the context of the 
phenomenological MSSM (pMSSM) with the neutralino as the lightest SUSY particle (LSP)~\cite{Djouadi:1998di}. 
We perform this study taking the mass of the heavy pseudoscalar, $M_A$, and the ratio of the vacuum 
expectation value of the two Higgs doublets, $\tan \beta$, as the main parameters. We highlight the 
complex pattern of decays arising at low values of $\tan \beta$, values which are shown to be compatible 
with the present data and discuss the complementarity of decay modes such as $H \rightarrow ZZ$, $tt$ 
and $hh$ and $A \rightarrow Zh$. 

The combination of the relevant decay channels to extend the sensitivity of the heavy Higgs searches 
over most of the $[M_A - \tan \beta]$ was already discussed in~\cite{ATLASTDR}. Here, we use the published 
and preliminary results for the expected upper limits on the product of production cross section and decay 
branching fraction in several channels as constraints and extrapolate them to the full 2012 data set of 
25~fb$^{-1}$/experiment at 8~TeV and to 150~fb$^{-1}$ of 14~TeV data. 
 
In section II, we discuss the production and decays of the $H$ and $A$ neutral bosons, with special emphasis 
for the low $\tan \beta$ region. 
Section  III is devoted to presentation of the results of our systematic study of the indirect constraints 
derived from the latest measurements of the decay rates of the 126~GeV Higgs-like particle together with those 
obtained from direct searches for $H$/$A \rightarrow \tau^+ \tau^-$, $H_{SM} \rightarrow ZZ$, $bbH \rightarrow bbbb$ 
and resonant $tt$ production and their expected sensitivity on the 14~TeV LHC data. Then, we review additional 
decay modes, which have not yet been considered in the LHC searches but will become important at 14~TeV in the low 
$\tan \beta$ region and characterise their kinematics and reconstruction strategies. Finally, we 
discuss the validity of these bounds when taking into account the production cross section uncertainties and 
the role of SUSY particles affecting the decays of heavy Higgs bosons, either in their direct
decays to SUSY states or through loop corrections to their decay widths. Section IV has the conclusions.

\section{The Higgs Sector and the $\pmb{M_A - \tan\beta}$ Parameters}

\subsection{$H$ and $A$ production and decays in the pMSSM}

The MSSM neutral heavy Higgs bosons $H$ and $A$ have couplings modified compared to the SM Higgs state. 
In the decoupling limit ($M_A \gg M_Z$), the $H/A$ coupling to the top quarks is suppressed by $1/\tan\beta$, 
while the couplings to bottom quarks and tau leptons are enhanced by $\tan\beta$. As a consequence, 
the $H/A tt$ coupling is important only for $\tan\beta\lesssim 10$, those to $bb$ and $\tau\tau$ becoming 
dominant for larger values.
On the other hand, the $H$ couplings to vector bosons are suppressed by a factor $\cos(\beta-\alpha)$, which 
in the large $M_A$ and $\tan\beta$ limit, decreases as $1/\tan\beta$. The situation is the same for the $AhZ$ 
coupling, while there is no $A$ coupling to vector bosons at tree level.
Finally, the coupling of the $H$ to $hh$ also decreases in the large $M_A$ and $\tan\beta$ limit with 
$1/\tan\beta$. Hence, the description of the heavy Higgs sector in the large $\tan\beta$ limit is simply 
dominated by the couplings to $b$ and $\tau$ fermions, whereas in the small $\tan\beta$ regime, a rich 
phenomenology emerges, as the other couplings become important. A thorough discussion can be found in 
Ref.~\cite{Djouadi:2005gj}.

% The branching fractions of heavy Higgs bosons in the MSSM have been discussed in details in 
% Ref.~\cite{Djouadi:2005gj}.
% We review here the features most relevant to the extraction of the constraints in the 
% $[M_A - \tan \beta]$ plane from the LHC results. 

\begin{figure}[t!]
\begin{tabular}{cc}
\includegraphics[width=0.5\columnwidth]{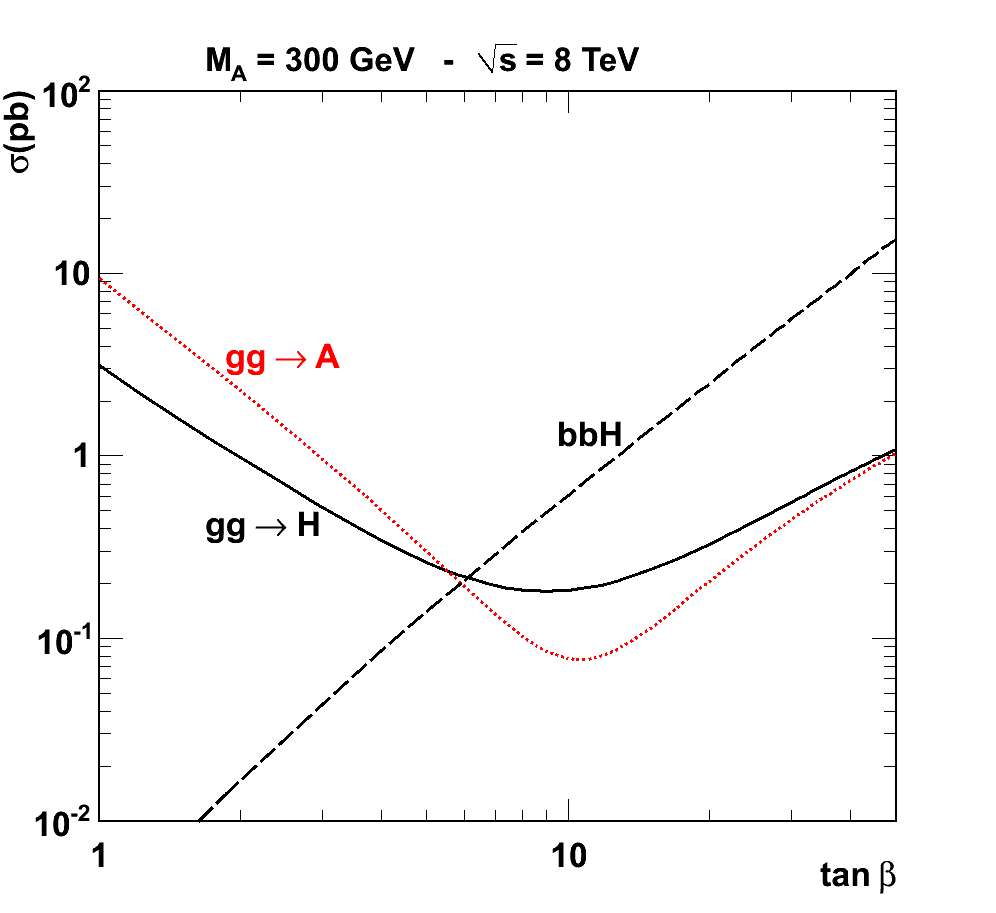} &
\includegraphics[width=0.5\columnwidth]{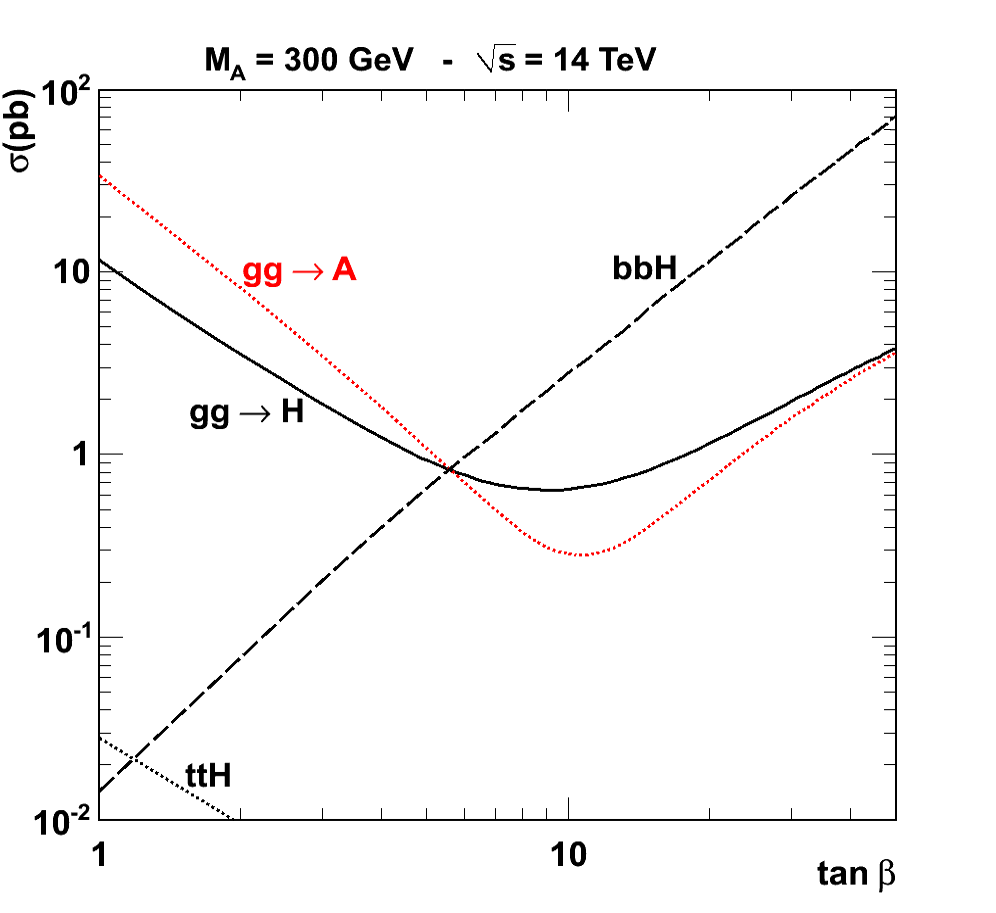} \\
\includegraphics[width=0.5\columnwidth]{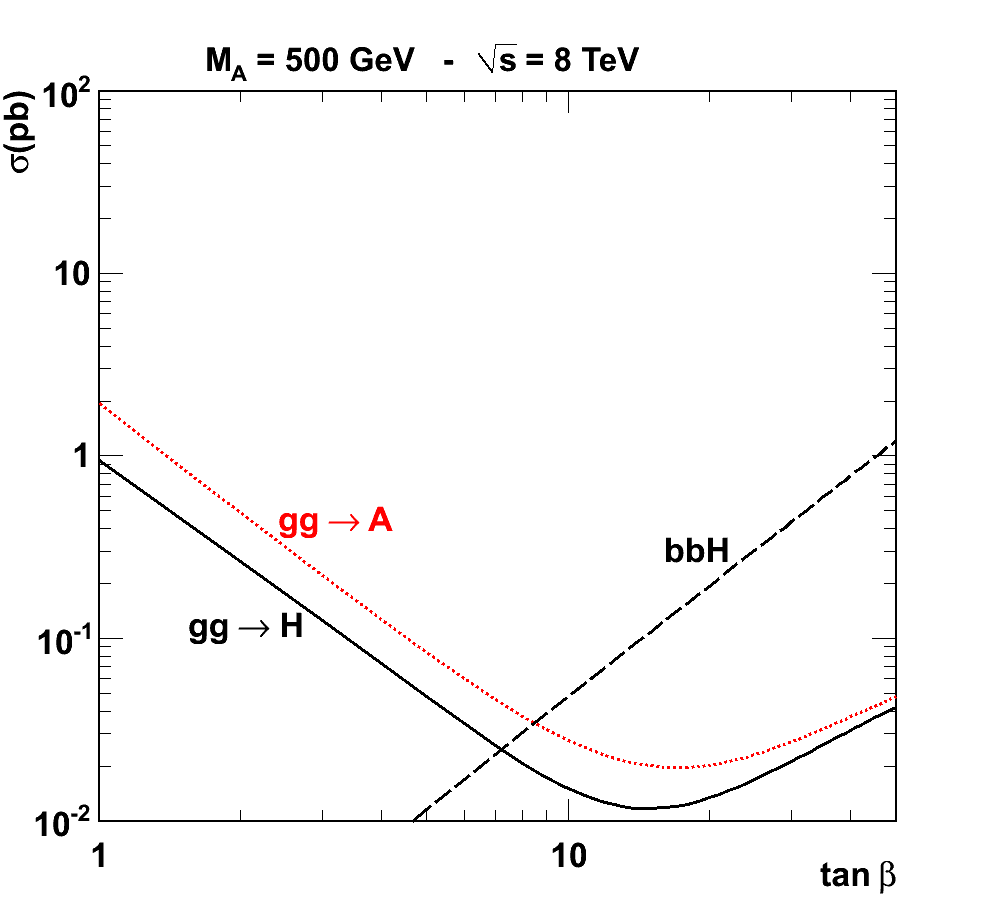} &
\includegraphics[width=0.5\columnwidth]{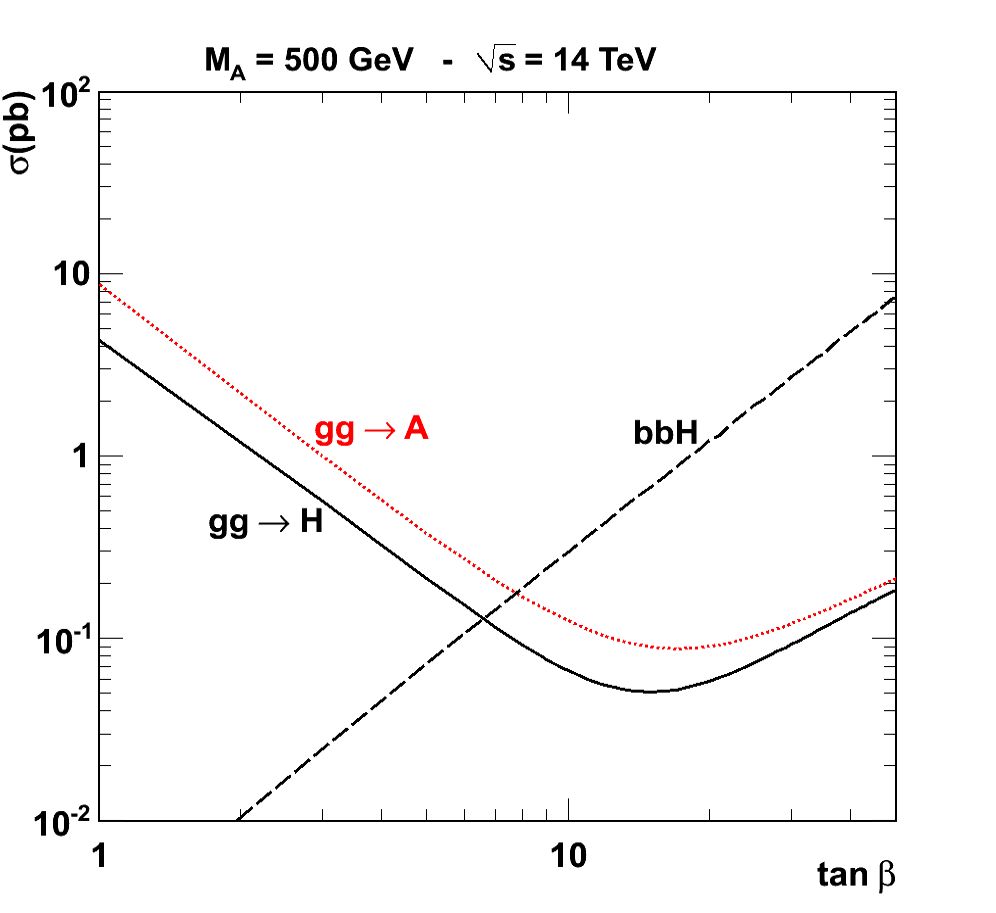} \\
\end{tabular}
\vspace*{-0.40cm}
\caption{Production cross sections for $H$ and $A$ bosons as a function of $\tan \beta$ in $pp$ 
collisions at 8~TeV (right panels) and 14~TeV (left panels) for $M_A$ = 300~GeV (upper panels) and 
500~GeV (lower panels).}
\label{fig:HAxs}
\end{figure}

The $H$ and $A$ production cross section is dominated by the gluon fusion process and the associate Higgs production 
with $b$ quarks. The relevant cross sections are shown in Fig.~\ref{fig:HAxs} for two values of the pseudoscalar Higgs 
mass ($M_A = $ 300 and 500 GeV) at $\sqrt{s}=$ 8 and 14~TeV, as a function of $\tan \beta$.

% For even smaller values of $\tan \beta$ the cross sections for the dominant 
% processes become even larger, with the exception of $H/A+bb$, as the $H$ couplings to top quarks and gauge bosons 
% are less suppressed. 

The $bbH$ associate production is a tree level process, which increases as $\tan^2\beta$ and becomes dominant for 
$\tan\beta\gtrsim 10$. Instead, the gluon fusion processes~\cite{Muhlleitner:2010zz}, induced by top and bottom quark 
loops, have the top loops dominant at small $\tan\beta$, resulting in a decrease of the total cross section with 
$\tan \beta$ up to the point where the $b$ loops take over and the total cross section increases. Finally, the 
$ttH$ production mode is kinematically suppressed and decreases with $1/\tan^2\beta$, while vector boson fusion and 
associate production with gauge bosons is not important, contrary to the case for the lightest Higgs boson.

\begin{figure}[t!]
\vspace*{-0.35cm}
\begin{tabular}{cc}
\includegraphics[width=0.5\columnwidth]{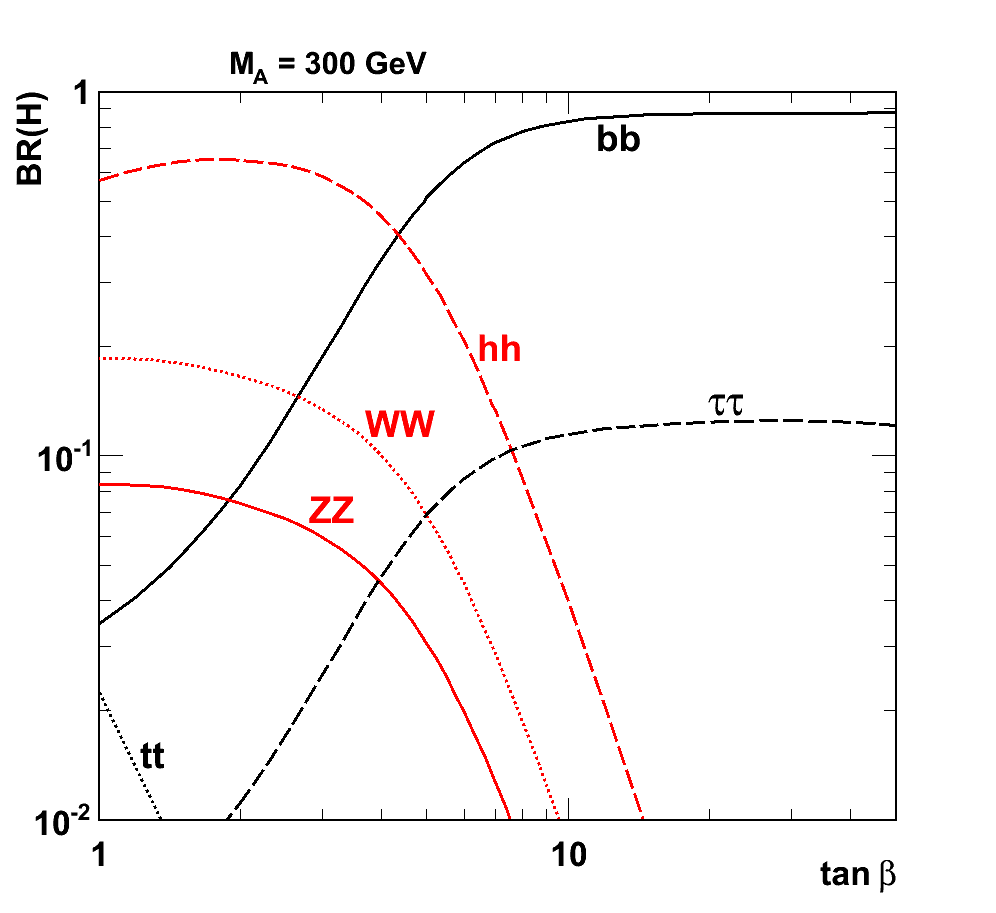} &
\includegraphics[width=0.5\columnwidth]{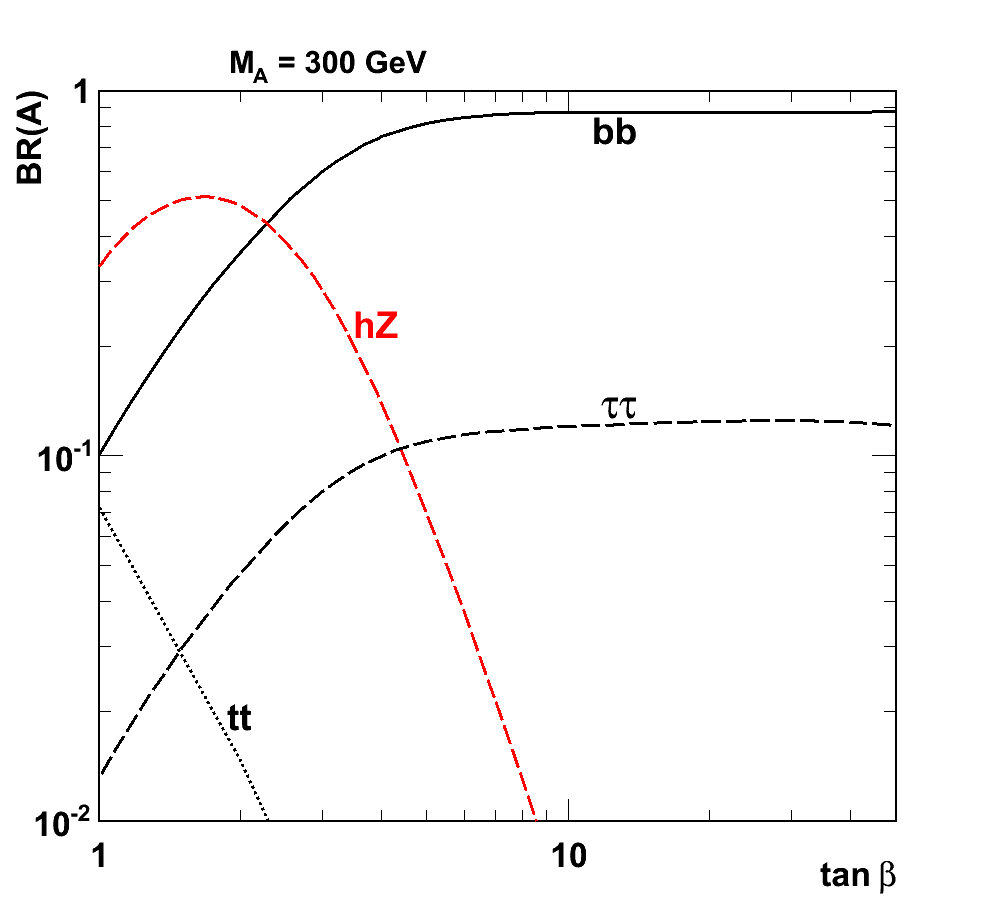} \\
\includegraphics[width=0.5\columnwidth]{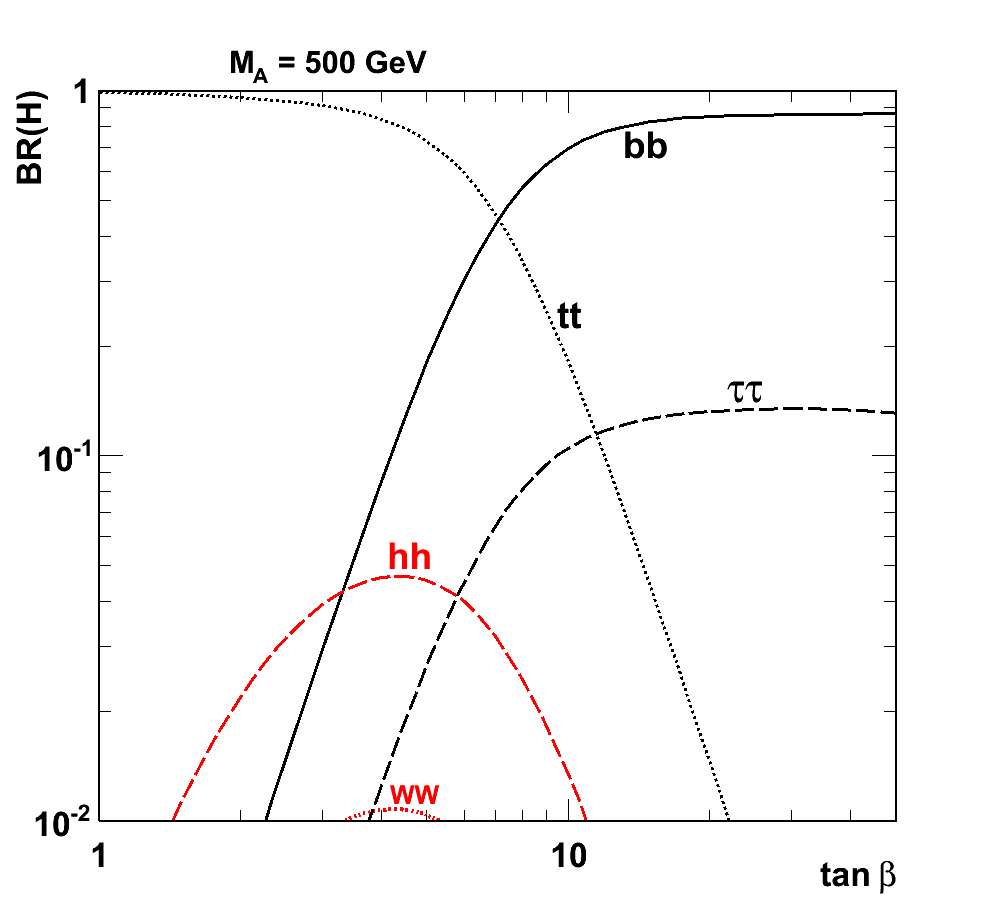} &
\includegraphics[width=0.5\columnwidth]{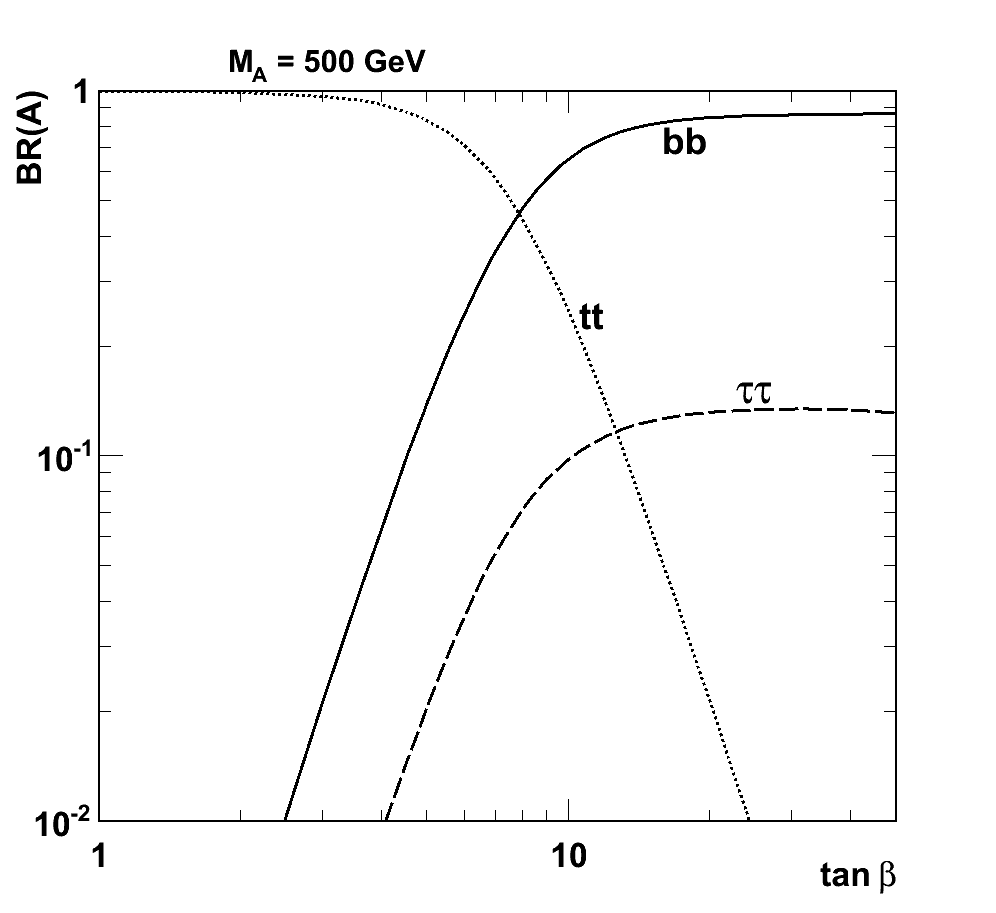} \\
\end{tabular}
\vspace*{-0.40cm}
\caption{Decay branching fractions for $H$ (left) and $A$ (right) bosons of mass 300~GeV (above) and 500~GeV (below) 
as a function of $\tan \beta$, in absence of decay channels into SUSY particles.}
\label{fig:HAbrTb}
\end{figure}

\begin{figure*}[ht!]
\begin{tabular}{ccc}
\hspace*{-0.25cm} \includegraphics[width=0.75\columnwidth]{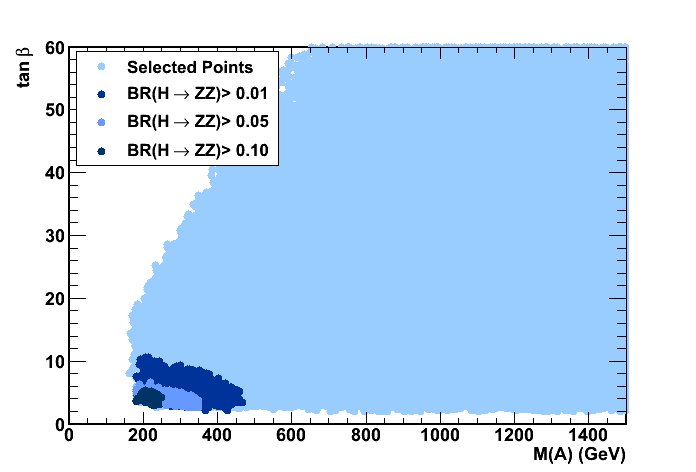} &
\hspace*{-0.75cm} \includegraphics[width=0.75\columnwidth]{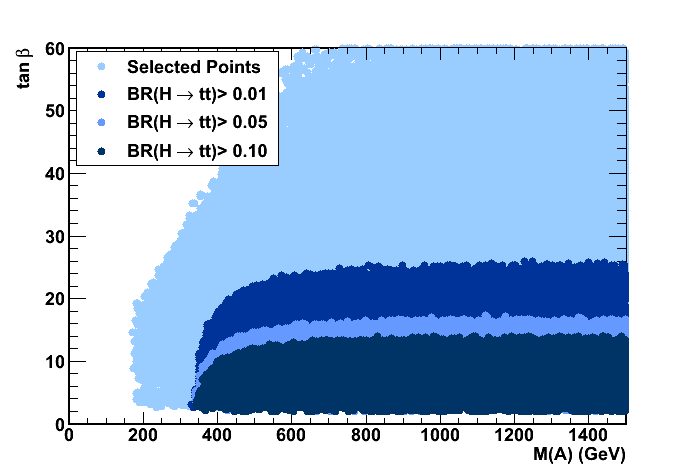} &
\hspace*{-0.75cm} \includegraphics[width=0.75\columnwidth]{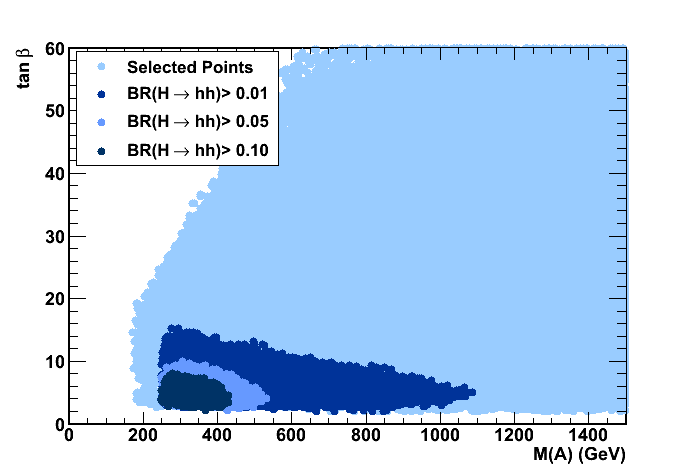} \\
\end{tabular}
\vspace*{-0.40cm}
\caption{Branching fraction for $H$ %and $A$ 
$\rightarrow ZZ$ (left), $\rightarrow t \bar t$ (centre) and 
$\rightarrow hh$ (right) for the selected pMSSM points in the $[M_A - \tan \beta]$ plane.}
\label{fig:MATbBR}
\end{figure*}
The decay $A$/$H \rightarrow \tau^+ \tau^-$ is the main process for the LHC experiments to 
search for the neutral heavy Higgs bosons at the present LHC energy, the dominant decay into $b \bar b$ 
being overwhelmed by the SM multi-jet background. As such, the $\tau \tau$ mode has so far attracted most 
of the attention in the LHC searches for heavy Higgs bosons. At intermediate to large values of $\tan \beta$ 
the $\tau \tau$ and $bb$ channels saturate the decay widths of the $A$ and $H$. 
%These couplings, proportional to $\tan \beta$, decrease for low $\tan \beta$ values, while those to $tt$, 
%scaling as $1/\tan\beta$ above threshold, increase. 
At low $\tan \beta$ the decay pattern of the heavier MSSM Higgs particles becomes more complicated by the onset of 
several decay modes which compete with $\tau \tau$, in particular $WW$, $ZZ$, $tt$ and $hh$.
The branching fractions for the decays of $H$ and $A$ bosons are shown in Fig.~\ref{fig:HAbrTb} as a function of 
$\tan \beta$ for two masses below (300~GeV) and above (500~GeV) the $t \bar{t}$ threshold. 

The main features can be summarised as follows. Below the $tt$ threshold, the $H$ boson decays into gauge bosons 
$H \to WW, ZZ$ and into pairs of the light Higgs boson, $hh$ have substantial rates. In the interval 
$2M_h \lsim M_H \lsim 2m_t$, this interesting channel, $H\to hh$, becomes the dominant decay mode for $\tan \beta \sim 3$.
Similarly the pseudoscalar $A$ boson decays into $hZ$ with a significant  rate, above threshold and at low 
$\tan \beta$. For heavier $H/A$ masses, the top decay channels, $H, A \rightarrow t\bar{t}$, becomes dominant below 
$\tan \beta \sim 5-10$. 

Figure~\ref{fig:MATbBR} shows the regions in the $[M_A - \tan \beta]$ plane of the pMSSM parameter space, where 
the branching fractions of $H \to ZZ/WW$, $H/A \to t\bar t$ and $H \to hh$ are larger than 1\%, 5\% and 10\%. 
As can be seen, BR($H/A \to ZZ/WW, hh$) can be large in the small to intermediate $\tan \beta$ and $M_A$ region. 
Above the threshold, the BR($H \to t\bar t$) is large for $\tan \beta \lesssim 20$ independently of the $M_A$ value.

The exclusion limits in the $[M_A - \tan \beta]$ plane may be modified when some light SUSY particles are present in 
the spectrum. In particular, three scenarios can affect the $[M_A - \tan \beta]$ bounds. First, light SUSY particles, 
with  masses $ \lsim \frac12  M_{H/A}$ may induce SUSY decays of the $H/A$ states thus reducing the 
$H/A \to \tau^+\tau^-$ branching fraction. 
For $M_A \lsim 1$~TeV, these SUSY particles can be light neutralinos or charginos, indicated collectively with 
$\tilde{\chi}$ in the following, and light sleptons, in particular staus, $\tilde{\tau}$, in the case of the CP--even 
$H$ boson while for the pseudoscalar $A$, only decays $A \to \tilde \tau_1 \tilde \tau_2$ are allowed. 

The pMSSM, due to the uncorrelated mass values of the SUSY particles afforded by its 19 free parameters, 
offers a convenient framework for this study, in particular by revealing scenarios where decays into SUSY 
particles may be important. The scans used for this study with the constraints and the 
relevant ranges for the variation of the pMSSM parameters have been already presented in 
Ref.~\cite{Arbey:2012dq,Arbey:2012bp}. 
In this analysis, we start from a large sample of 2$\times$10$^8$ generated pMSSM points and select those 
fulfilling the constraints from LEP data, flavour physics, dark matter and $\tilde g$, $\tilde q$ direct 
searches at LHC as discussed in Ref.~\cite{Arbey:2012dq}. In particular, we apply the constraints derived from 
the rare decay $B_s \rightarrow \mu \mu$ and direct dark matter searches, also providing us with constraints 
to the $[M_A - \tan \beta]$ parameter space~\cite{Arbey:2011aa,Arbey:2012ax}, and we impose that one of the 
neutral Higgs bosons has a mass in the range 121.5 to 129.9~GeV to be consistent with the results of the 
SM Higgs searches, as discussed below. 

The LHCb experiment has recently announced the first evidence for the $B_s\to\mu^+\mu^-$ decay and measured its branching 
fraction to be in agreement with the SM expectation~\cite{Aaij:2012nna}. This branching ratio is sensitive to the Higgs 
sector, in particular to $M_A$ and $\tan\beta$, proportional to $\sim\tan^6\beta/M_A^4$ in the large $\tan\beta$ limit. 
Complementary information is also obtained by dark matter direct detection experiments, in particular the latest XENON-100 
limits~\cite{Aprile:2012nq}, probing the scattering of neutralino with matter, which can be mediated by scalar particles.

The tools used to perform the scans and the analysis have been presented in Ref.~\cite{Arbey:2011un,Arbey:2011aa}. 
Most relevant to this study are the calculations of the Higgs decay branching fractions and production cross sections.
The first are computed using the latest version of {\tt HDECAY (5.10)}~\cite{Djouadi:1997yw}. The cross section for 
$gg \rightarrow H/A$ process is computed at NNLO with {\tt HIGLU 3.1}~\cite{Spira:1995rr,Spira:1996if},  
that  for $bb \rightarrow H/A$ at NNLO with {\tt bbh@nnlo}~\cite{bbh} and that for $pp \rightarrow bbH$ at 
LO with {\tt HQQ}~\cite{Spira:1997dg}.
In addition, we compare the results for $gg$ and $bb \rightarrow H/A$ from these programs to those from 
{\tt SusHi}~\cite{Harlander:2012pb} and found an agreement within 10-15\%. The Higgs and superparticle spectra 
are calculated with {\tt Softsusy 3.2.3}~\cite{softsusy} and {\tt SuperIso Relic v3.2}~\cite{superiso,superiso_relic}
computing the dark matter relic density and flavour constraints and providing the central control program 
interfaced to the other codes.

\subsection{SUSY Effects in $H$ and $A$ Decays}

There are regions of the MSSM parameter space where the $\tau \tau$ channel is suppressed and the limits derived in this 
channel are correspondingly relaxed. These may be due to direct decays of $H$/$A$ to SUSY particles or to loop corrections 
to the $H/A bb$ vertices, affecting the $H/A \to \tau \tau$ branching fraction.

We consider first the decays of heavy neutral Higgs bosons into pairs of SUSY particles. 
The heavy Higgs bosons couple to charginos and neutralinos, primarily to identical particles for the mixed gaugino/higgsino states, 
and to different particles in case of pure gaugino or higgsino states. If the decay to charginos is allowed, it dominates over the 
decays to neutralinos. Heavy neutral Higgs bosons also couple to scalar fermions. However, decays to scalar fermions of the first two 
generations are suppressed and only significant at low $\tan\beta$, where they are sub-dominant. For scalar fermions of the third 
generation the decay rates can be much larger, but they are suppressed at large $\tan\beta$ for scalar top quarks, while they are 
enhanced for the scalar taus and scalar bottoms. Since the lightest scalar tau, $\tilde{\tau}_1$ is often the NLSP at large 
$\tan\beta$, decays to staus are usually the dominant channel for decays into scalar fermions.

\begin{figure}[h!]
\vspace*{-0.25cm}
 \begin{tabular}{cc}
 \includegraphics[width=0.5\columnwidth]{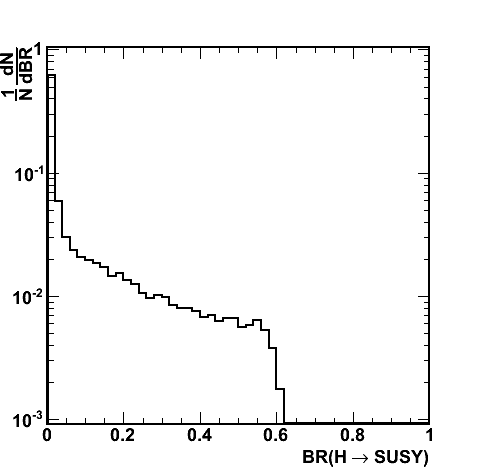} &
 \includegraphics[width=0.5\columnwidth]{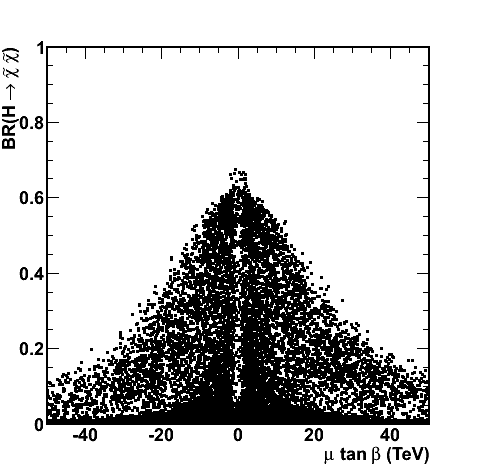} \\
 \end{tabular}
\vspace*{-0.40cm}
\caption{Decay branching fractions for $H$ into SUSY particles. In the left panel branching fraction for inclusive decays into any 
pairs of SUSY particles for the accepted pMSSM points where at least one of these decay channels is kinematically allowed. 
In the right panel, branching fraction into pairs of charginos and neutralinos, $\tilde{\chi} \tilde{\chi}$ as a function of 
$\mu \tan \beta$.}
\label{fig:BRHSUSY}
\end{figure}

Figure~\ref{fig:BRHSUSY} shows the decay branching fraction of $H$ into any pair of SUSY particles calculated for the accepted 
pMSSM points for which at least one of these decay channels is kinematically allowed. In approximately 25\% of these cases the 
branching fraction into SUSY particles is larger than 0.10.
%The $\chi \chi$ represents the sum of all the kinematically accessible chargino and neutralino pairs and 
%$\tilde{\tau} \tilde{\tau}$ represents the sum of 
%$\tilde{\tau}_1 \tilde{\tau}_1$, $\tilde{\tau}_1 \tilde{\tau}_2$ and $\tilde{\tau}_2 \tilde{\tau}_2$ for the $H$ and 
%the $\tilde{\tau}_1 \tilde{\tau}_2$ final state for the $A$ boson. 

\begin{figure}[h!]
\vspace*{-0.25cm}
\includegraphics[width=0.75\columnwidth]{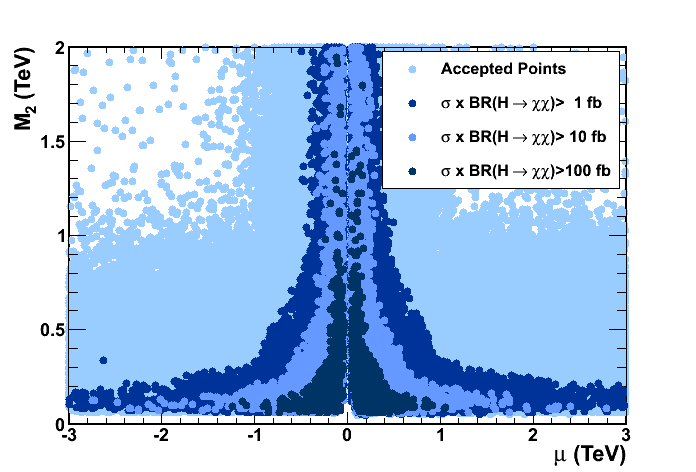} \\
\vspace*{-0.40cm}
\caption{Product of production cross section and decay branching fraction for $H \rightarrow \tilde{\chi} \tilde{\chi}$ 
at 14 TeV in the $[\mu - M_2]$ parameter plane.  The dots in the light colour show all the selected pMSSM points and those 
in darker shades of colour the points having $\sigma \times \mathrm{BR}$ larger than 1, 10 and 100~fb.}
\label{fig:XSBR14MuM2}
\end{figure}
The yield in the $\tilde{\chi} \tilde{\chi}$ channels,  representing the sum of all the kinematically accessible chargino 
and neutralino pairs, depends on the mass parameter $M_2$ and the Higgsino mass mixing parameter $\mu$. 
Figure~\ref{fig:XSBR14MuM2} shows the $\sigma \times \mathrm{BR}$ product in the $[\mu - M_2]$ parameter plane to 
highlight the enhancement of this class of decays along the small $M_2$ or $\mu$ regions.

The rates of decays into SUSY particles depend mostly on the difference between the masses of heavy boson and those of the 
SUSY particles. As the scale of the mass of the $H$ and $A$ bosons probed at the LHC increases, decays into SUSY particles 
become more likely and have to be carefully considered. The relevant mass patterns are extensively probed in our pMSSM scans.  
The increase in the branching fractions of any of these SUSY channels is correlated to the decrease of 
that for the $\tau \tau$ mode, which can be suppressed by a factor of two, or more, compared to its average value 
at large $M_A$ and $\tan \beta$ values.
%Figure~\ref{fig:H0tau} shows the regions of the $[M_A - \tan \beta]$ plane 
%where the BR($H \rightarrow \tau^+ \tau^-$) drops below 0.07. Since the mean value of the 
%branching fraction at large $M_A$ and $\tan \beta$ values is $\sim$ 0.13, these regions correspond to 
%a suppression of more than 50\%.

\begin{figure}[h!]
\vspace*{-0.35cm}
\begin{center}
 \begin{tabular}{cc}
 \includegraphics[width=0.5\columnwidth]{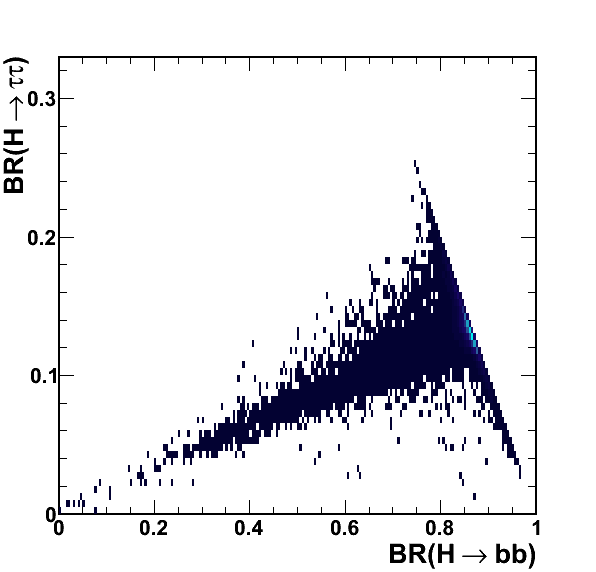} &
 \includegraphics[width=0.5\columnwidth]{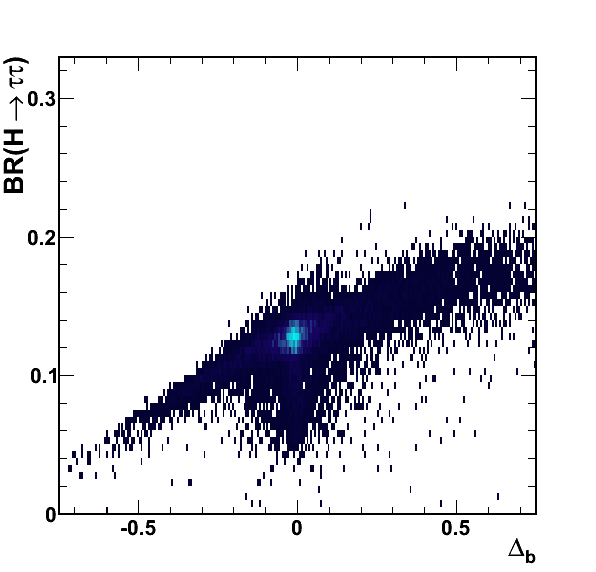} \\
 \end{tabular}
\end{center}
\vspace*{-0.40cm}
\caption{Correlation of the BR($H \rightarrow \tau \tau$) with BR($H \rightarrow b \bar b$) (left) and with $\Delta_b$ 
(right) for accepted pMSSM points. The correlated suppression of both the $\tau \tau$ and $b \bar b$ branching fractions 
are due to additional decays into SUSY particles, while the decrease of $\tau \tau$ with the increase of $bb$ is due to 
$\Delta_b$ effect.}
\label{fig:BRHBRh}
\end{figure}

Finally loop corrections to the $Hbb$ and $Abb$ vertices, known as $\Delta_b$ corrections~\cite{Pierce:1996zz}, 
modify both the $H/A$ production rates and their decay widths. In the decoupling limit, the $H/A$ coupling to $bb$ is 
modified by a factor $(1+\Delta_b)^{-1}$, where
\begin{eqnarray}
\Delta_b &\approx& \frac{2\alpha_s}{3\pi}\, \mu M_3 \,\frac{\tan\beta}{{\rm max}(M_3^2,m_{\tilde{b}_1}^2,m_{\tilde{b}_2}^2)} \nonumber\\
&&+ \frac{\mu A_t}{16\pi^2}\, \frac{y_t^2 \tan\beta}{{\rm max}(\mu^2,m_{\tilde{t}_1}^2,m_{\tilde{t}_2}^2)}.
\end{eqnarray}%
Full one loop corrections to the $WW$ and $ZZ$ decays have also been computed~\cite{Hollik:2011xd,Gonzalez:2012mq}. 
We observe that the BR($H/A \rightarrow \tau \tau$) is reduced as a result of the enhancement of BR($H/A \to b \bar b$) 
due to these corrections for SUSY parameters yielding a large $\Delta_b$ of negative sign (see Figure~\ref{fig:BRHBRh}).  
Such a large, negative $\Delta_b$ term has also implications on the decay branching fractions of the lightest $h$ 
boson. In the region where the $H \rightarrow \tau \tau$ decay rate is reduced, the branching fraction BR($h \rightarrow b \bar b$) 
is also reduced and those for the other modes correspondingly increased. These patterns might be tested through more precise 
determinations of the signal strengths of the lightest Higgs decays.
In view of these effects, redundancy obtained through search in 
multiple channels sensitive in the same regions of the $[M_A - \tan \beta]$ parameter space appears to be essential.

\subsection{Constraints from the $h$ Mass and Decay Rates}

Assuming that the observed $\sim$126~GeV state is the lightest Higgs boson of the MSSM, $h$, its mass $M_h$ depends
on several SUSY parameters, in particular $M_A$, $\tan \beta$ and the SUSY scale, $M_S = \sqrt{m_{\tilde t_1} m_{\tilde t_2}}$. 
\begin{figure}[hb!]
\vspace*{-0.35cm}
\includegraphics[width=0.55\columnwidth]{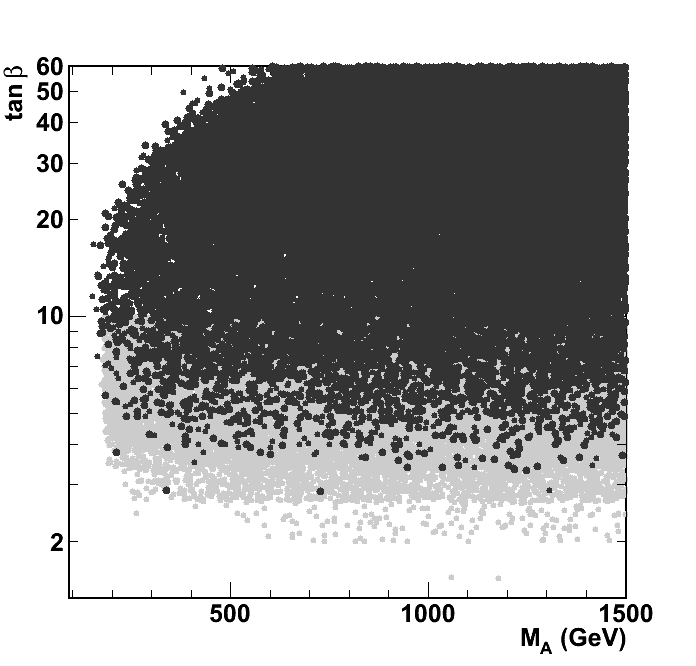}
\vspace*{-0.40cm}
\caption{The $[M_A - \tan \beta]$ parameter space compatible with 121.5 $<M_h<$ 129.9~GeV for different SUSY scales $M_S$. 
Distribution of the accepted pMSSM points in the $[M_A - \tan \beta]$ compatible with the $M_h$ mass interval for 
0.5$< M_S <$3.5~TeV (black dots) and 5$< M_S <$20~TeV (light grey dots).}
\label{fig:MSTb}
\end{figure}
The LEP-2 limit~\cite{Barate:2003sz}, $M_h>$ 114.5~GeV has been long used to define a constraint in the $[M_A - \tan \beta]$ plane, 
corresponding to $\tan \beta \gsim$2.4 for $M_S$ = 1~TeV and $M_{\mathrm{top}}$ = 172.9~GeV~\cite{Schael:2006cr} in the so-called 
$M_h^{max}$ scenario~\cite{Carena:2002qg}. 
Now, each value of $M_h$ defines a constraint in the 
$[M_A - \tan \beta]$ which depends on $M_S$. A larger value of $M_S$ corresponds to a weaker constraint on $\tan \beta$. 
Therefore, it is possible to set a large enough $M_S$ scale which recovers the low $\tan \beta$ solutions of the MSSM, even by 
applying the LHC constraints for $M_h$, which are now significantly stronger than the LEP-2 limit.  This is illustrated in 
Figure~\ref{fig:MSTb} which shows the values of $\tan \beta$ vs.\ $M_A$ for our pMSSM scans which are compatible with 
121.5$< M_h <$129.9~GeV, for two intervals of values of the SUSY scale $M_S$.  
Large enough $M_S$ values rescue the MSSM scenarios at low values of $\tan \beta$, provided we accept a high fine tuning parameter 
from the large scale of $M_S$. This observation motivates the special attention we have chosen to devote to low $\tan \beta$ 
scenarios in this study.

\begin{figure}[hb!]
\vspace*{-0.20cm}
\begin{tabular}{cc}
\hspace*{-0.25cm} \includegraphics[width=0.52\columnwidth]{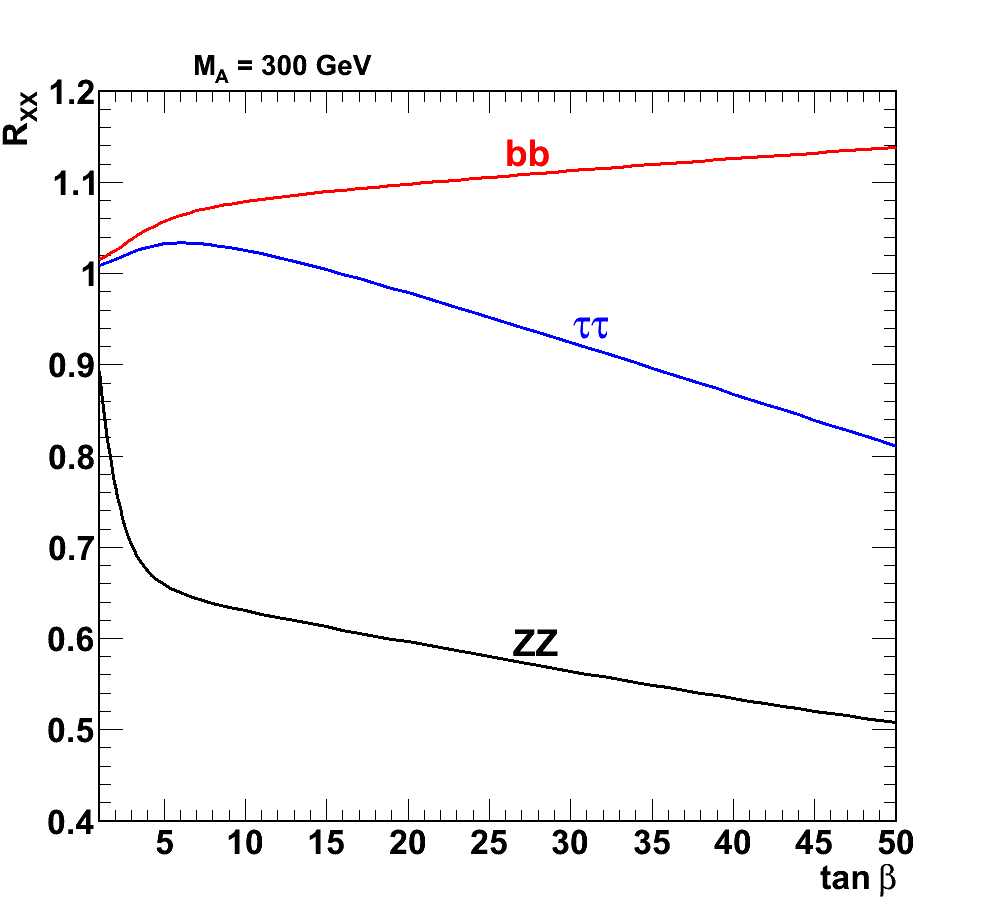} &
\hspace*{-0.35cm} \includegraphics[width=0.52\columnwidth]{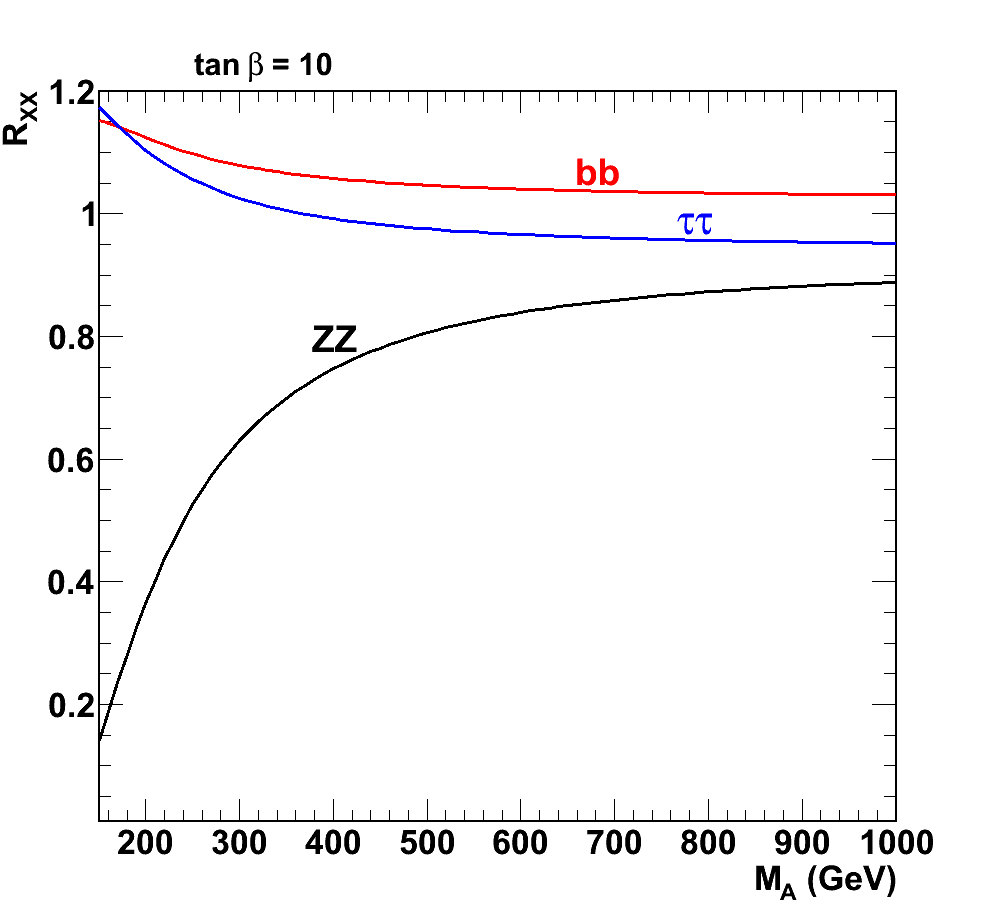} \\
\end{tabular}
\vspace*{-0.40cm}
\caption{Scaling of the $h$ branching fractions into $b \bar b$, $\tau \tau$ and $ZZ$ normalised to their SM value 
as a function of $\tan \beta$ for $M_A$ = 300~GeV (left panel) and as a function of $M_A$ for $\tan \beta$ = 10 (right panel).}
\label{fig:BRh}
\end{figure}
A second set of indirect constraints is derived by the measured $h$ decay rates. For large $M_A$ values, the couplings of 
the $h$ boson can be expanded in powers of $M_Z/M_A$ to obtain the following tree-level result~\cite{Djouadi:2005gj}:
\beq
g_{hVV} \stackrel{\small M_A \gg M_Z} \lra
1-\frac{M_Z^4}{8M_A^4}\sin^2 4\beta \stackrel{\small \tb \gg 1} \lra 
1-\frac{2 M_Z^4}{M_A^4\tan^2 \beta}
\label{gHVVdecoup}
\eeq
For $M_A\gg M_Z$,  $g_{hVV}$ reaches the SM value, more quickly if $\tb$ is large.
The $h$ couplings to up- and down-type fermions scale as~\cite{Djouadi:2005gj}:
\beq
\label{gHff:decoup}
g_{huu}\stackrel{\small M_A \gg M_Z} \lra 1+\frac{M_Z^2}{2 M_A^2} 
\frac{\sin 4\beta}{\tan \beta} \stackrel{\small \tb  \gg 1} \lra \    
1-\frac{2 M_Z^2}{M_A^2 \tan^2 \beta}\\ 
g_{hdd}\stackrel{\small M_A \gg M_Z} \lra \ 1-\frac{M_Z^2}{2 M_A^2} 
\sin 4\beta \tan \beta \ \stackrel{\small \tb  \gg 1} \lra \    
1+\frac{2 M_Z^2}{M_A^2}
\eeq 
The couplings of the $h$ boson approach those of the SM Higgs boson for $M_A \gg M_Z$ and these limits are reached 
at lower values of $M_A$ for large $\tb$ (see Figure~\ref{fig:BRh}). 
In practice, the ratio of branching fractions 
$R_{XX} = {\mathrm{BR}}(h \rightarrow XX) / {\mathrm{BR}}(H_{\mathrm{SM}} \rightarrow XX)$ or the signal strengths 
$\mu_{XX} = \sigma (h)/ \sigma (H_{\rm SM}) \times R_{XX}$, where $\sigma$ is the relevant production cross section, 
can be used to set constraints on the value of $M_A$. The recent approximate N$^3$LO calculation of the Higgs production 
cross section resulting in a 17\% correction also needs to be taken into account~\cite{Ball:2013bra}.
The latest set of LHC results already allows us to evaluate 
some non-trivial constraints, as discussed in the next section.

\section{Constraints in the $\pmb{M_A - \tan \beta}$ plane}

The LHC searches have gathered a significant corpus of results, which can be used to place some important 
constraints on the $H$ and $A$ bosons in a variety of channels. These results also allow us to study the expected 
sensitivity of data to be taken at 13 and 14~TeV from 2015. In the next two sections we discuss the current constraints 
and in the following we present the extrapolation to 14~TeV. There are important decay channels, such as 
$H/A \rightarrow hh$ and $A \rightarrow hZ$, for which no analysis has been performed yet on the LHC data. 
We characterise the kinematics and reconstruction strategy for these processes using parametrised simulation at 
the end of this section.  

\subsection{Present Constraints (7 and 8~TeV)}

\subsubsection{Indirect Constraints from the light Higgs signal}

The ATLAS and CMS collaboration have recently updated their determination of the mass and signal strengths of the 
Higgs-like particle. In particular, results for the $\gamma \gamma$~\cite{ATLAS-CONF-2013-012,CMS-13-001}, 
$ZZ$~\cite{ATLAS-CONF-2013-013,CMS-13-002} and $WW$~\cite{ATLAS-CONF-2013-030,CMS-13-003} channels have been reported 
by both collaborations for the full 8~TeV data set, corresponding to integrated luminosities of up to 25~fb$^{-1}$. 
In addition, CMS has updated the search in the $\tau \tau$ channel at low mass~\cite{CMS-13-004}. 
Here we use the weighted averages for the mass and signal strengths in these preliminary results, as summarised in 
Table~\ref{tab:input}. For the important $\gamma \gamma$ channel we average the preliminary results of the multi-variate and 
cut-based analyses of CMS accounting for the quoted correlation~\cite{Schmelling}. The results of the two collaborations 
are only marginally consistent and we therefore rescale the error of the combined result according to the prescriptions of the 
Particle Data Group~\cite{Beringer:1900zz}. We use these new inputs and perform an analysis of the regions of MSSM parameter space 
favoured by these data. The analysis follows the strategy discussed in \cite{Arbey:2012bp}. We define the 90\% C.L. region corresponding 
to observables given in  Table~\ref{tab:input} by constructing the corresponding $\chi^2$ probability. We account for the 
theory uncertainties on the MSSM $h$ mass, $\pm$1.5~GeV, and the Higgs production rates, $\pm$20\%.
No signal evidence has been reported for the $b \bar b$, where we also include the combined estimate on $\mu_{bb}$ obtained by 
CDF and D0 at the Tevatron~\cite{Aaltonen:2012qt}, and the $\tau^+ \tau^-$ channels. For these we add the contribution 
to the total $\chi^2$ only when the respective $\mu$ value is outside the $\pm$1.5~$\sigma$ interval from the measured central value. 
Compared to the results available at the end of 2012, we register a marked realignment of the average values for the $\mu_{XX}$ 
signal strengths from the ATLAS and CMS results around the SM values. This has important consequences on the constraints derived. 
In particular, the $M_A$ bound derived from the new data is about 100~GeV lower compared to that obtained on the first preliminary 
results on part of the 8~TeV data released at the end of 2012, without the CMS re-analysis of the 
$\gamma \gamma$ channel~\cite{Arbey:2012bp}. This clearly shows that it is difficult to predict the sensitivity achievable in future 
for these indirect limits, which depends not only on the accuracy of the inputs but also rather critically on the measured values. 

\begin{table}[!h]
\begin{center}
\begin{tabular}{|c|c|c|}
\hline
Parameter & Value & Experiment \\ \hline \hline
$M_h$ (GeV)   & 125.7$\pm$0.4 & ATLAS\cite{ATLAS-CONF-2013-014}+CMS\cite{CMS-13-002} \\ 
$\mu_{\gamma \gamma}$ & 1.20$\pm$0.30 & ATLAS\cite{ATLAS-CONF-2013-012}+CMS\cite{CMS-13-001} \\
$\mu_{Z Z}$ & 1.10$\pm$0.22 & ATLAS\cite{ATLAS-CONF-2013-013}+CMS\cite{CMS-13-002} \\
$\mu_{W W}$ & 0.77$\pm$0.21 & ATLAS\cite{ATLAS-CONF-2013-030}+CMS\cite{CMS-13-003} \\                                 
\hline
$\mu_{b \bar b}$  & 1.12$\pm$0.45 & ATLAS\cite{ATLAS-2012-161}+CMS\cite{CMS-12-044}+(CDF+D0)\cite{Aaltonen:2012qt}\\ 
$\mu_{\tau \tau}$ & 1.01$\pm$0.36 & ATLAS\cite{ATLAS-2012-160}+CMS\cite{CMS-13-004}\\ 
\hline
\end{tabular}
\end{center}
\caption{Input values for the average values of the $h$ mass and signal strengths used for this study with their statistical 
accuracies. Systematic uncertainties are discussed in the text.}
\label{tab:input} 
\end{table}

\begin{figure}[hb!]
\vspace*{-0.20cm} 
\begin{center}
\begin{tabular}{cc}
\hspace*{-0.25cm} \includegraphics[width=0.54\columnwidth]{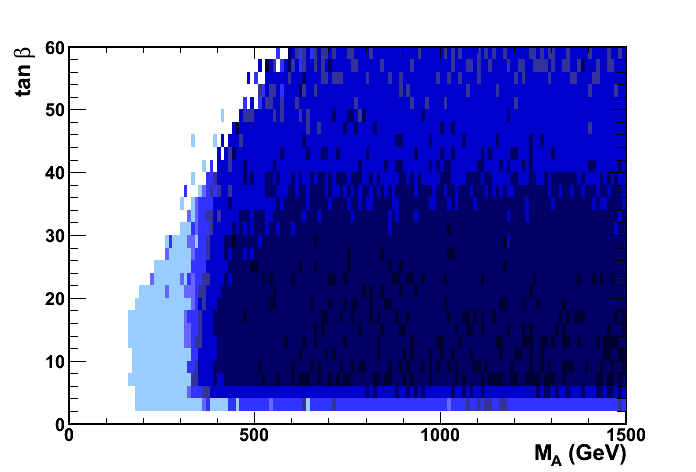} &
\hspace*{-0.65cm} \includegraphics[width=0.54\columnwidth]{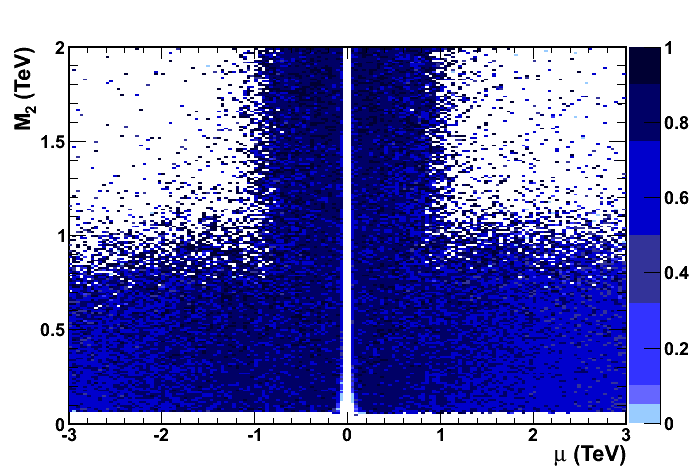} \\
\end{tabular}
\end{center}
\vspace*{-0.40cm}
\caption{Fractions of pMSSM points compatible at 90\% C.L. with the constraints of Table~\ref{tab:input} in the 
$[M_A - \tan \beta]$ (left) and $[\mu - M_2]$ (right).}
\label{fig:hMAtbRH}
\end{figure}

We consider points compatible at 90\% C.L. with these inputs accounting for theory uncertainties. We observe that these 
account for 76\% of the accepted pMSSM points, up from the 30\% obtained in the same analysis performed on the earlier data. 
For all these points the $\sim$126~GeV state observed by ATLAS and CMS is the lightest Higgs, $h$. Therefore we confirm 
the results from our previous analysis where we did not find any pMSSM solution compatible with the LHC Higgs results 
where the 126~GeV particle is either the $H$ or the $A$ boson. This result provides an answer to the question 
of~\cite{Bechtle:2012jw}. Figure~\ref{fig:hMAtbRH} shows this fraction 
as a function of $[M_A - \tan \beta]$ and $[\mu - M_2]$. From the $[M_A - \tan \beta]$ distribution of these points 
we define the region containing 99\% of the points compatible at 90\% C.L. with the LHC Higgs results. This region 
defines an indirect lower bound on $M_A$, which will be compared to the direct exclusion from $H$/$A$ searches in the 
next section.

\begin{figure*}[ht!]
\vspace*{-0.20cm}
\begin{tabular}{ccc}
\hspace*{-0.25cm} \includegraphics[width=0.75\columnwidth]{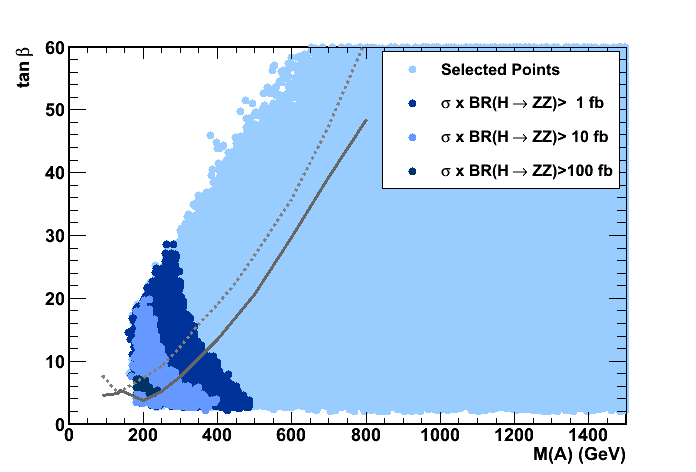} &
\hspace*{-0.75cm} \includegraphics[width=0.75\columnwidth]{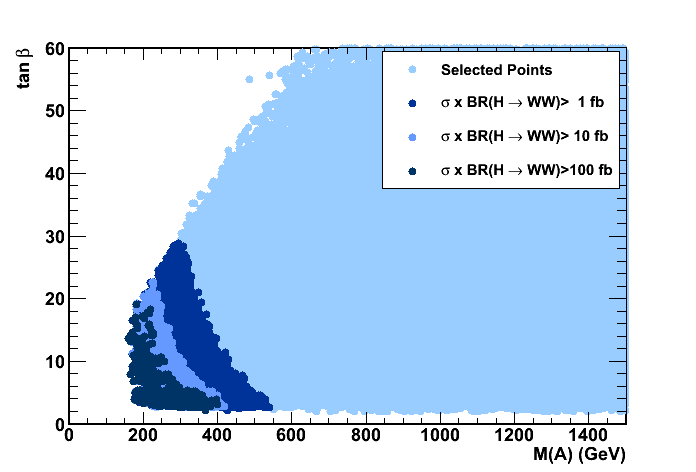} &
\hspace*{-0.75cm} \includegraphics[width=0.75\columnwidth]{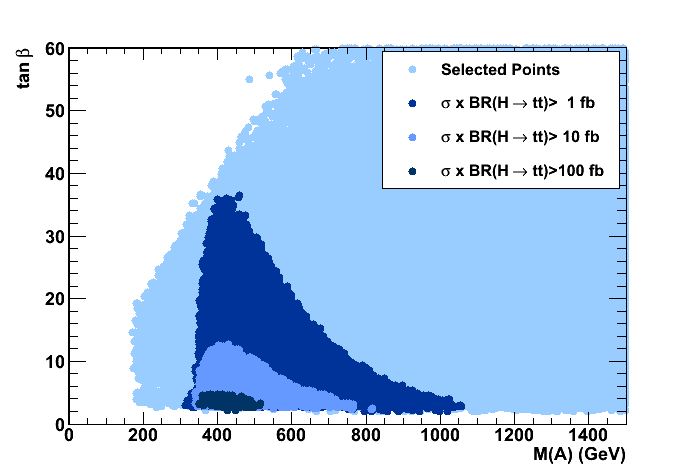} \\
\end{tabular}
\vspace*{-0.40cm}
\caption{Product of production cross section and decay branching fraction for $H \rightarrow ZZ$ (left), 
$H \rightarrow WW$ (upper centre) and $H \rightarrow t \bar{t}$ (right) at 8~TeV in the $[M_A - \tan \beta]$ parameter 
plane. The dots in the light colour show all the selected pMSSM points and those in darker shades of colour the points 
having $\sigma \times \mathrm{BR}$ larger than 1, 10 and 100~fb. The lines superimposed on the left panel show the expected 
(dashed) and observed (continuous) 95\% C.L. upper limits obtained in the $H/A \rightarrow \tau \tau$ search 
of~\cite{CMS-PAS-HIG-12-050}. Entries below threshold in the $t \bar t$ channel are due to off-shell decays.}
\label{fig:XSBR8}
\end{figure*} 
\subsubsection{Direct Constraints from MSSM Higgs Searches}

Searches for the  $H$/$A \rightarrow \tau^+ \tau^-$ process have been conducted by the ATLAS with 4.7 fb$^{-1}$ at 
7~TeV~\cite{Aad:2011rv} and CMS with 4.8+12.2 fb$^{-1}$ at 7 and 8~TeV~\cite{CMS-PAS-HIG-12-050}. The CMS sensitivity 
corresponds to an expected upper limit on the product of production cross section and decay branching fraction of 
$\sim$80~fb at 300~GeV and 20~fb at 500~GeV. In this study, we impose the expected CMS 
95\% C.L. limit on the product of production cross section and decay branching fraction, which is weaker than the 
observed limit, on our pMSSM points.

The production and decay pattern of the heavy MSSM neutral Higgs bosons crucially depend on the value of $\tan \beta$, as 
discussed above. The LHC data at 7 and 8 TeV, probe relatively large values, $\tan \beta \gsim$5--10. For these values, 
their couplings to $b$ quarks and $\tau$ leptons, proportional to  $\tan \beta$, are strongly 
enhanced, and those to top quarks and massive bosons, proportional to $\approx 1/\tan \beta$, are suppressed.
Therefore the $\tau \tau$ channel is presently the single most constraining decay mode. It defines a region of the 
$[M_A - \tan \beta]$ parameter space which is probed also by the $B_s \rightarrow \mu \mu$ rare decay and by dark matter 
direct detection experiments \cite{Arbey:2011aa,Arbey:2012ax}, but the $H/A \rightarrow \tau \tau$ LHC searches at 
7 and 8~TeV set the tightest constraints. 
In addition to it, preliminary results have been reported for the first search for 
$H$/$A \rightarrow b \bar{b}$ in associate production with $b$ jets  $bb H$/$A \rightarrow bbbb$ based on the 7~TeV 
CMS data~\cite{Behr:2013ji}, which has sensitivity at large values of $\tan \beta$ with an expected upper limit of 8~pb 
on $\sigma \times \mathrm{BR}$ at 300~GeV. The analyses of the decays of the SM Higgs $H_{SM} \rightarrow ZZ$ have set 
constraints on the product of production cross section and decay branching fraction 
$\sigma(gg \rightarrow H_{SM}) \times \mathrm{BR}(H_{SM} \rightarrow ZZ)$ for Higgs masses up 
to 1~TeV~\cite{ATLAS-CONF-2012-169,CMS-PAS-HIG-12-041}. 
These can now be used to constrain the decay of the heavy SUSY $H$/$A \rightarrow ZZ$, with upper limits of $\sim$1.9 
and 1.4~fb at 200 and 300~GeV, respectively.  Finally, the decay $H$/$A \rightarrow tt$ can be constrained 
through the cross section bounds obtained for the production of a narrow resonance decaying into top quark pairs, 
interpreted in the original studies in the context of the searches for the production of a lepto-phobic $Z'$ gauge boson, 
KK resonances or other exotic narrow resonances. Results have been reported by both ATLAS \cite{ATLAS:2012txa} and 
CMS~\cite{CMS:2012rq} for the 7~TeV data with cross section upper limits of order of 3~pb and 0.8~pb at resonance masses 
of $\sim$500~GeV and 800~GeV, respectively. 
\begin{figure}[h!]
\vspace*{-0.20cm} 
\begin{tabular}{c}
\includegraphics[width=0.975\columnwidth]{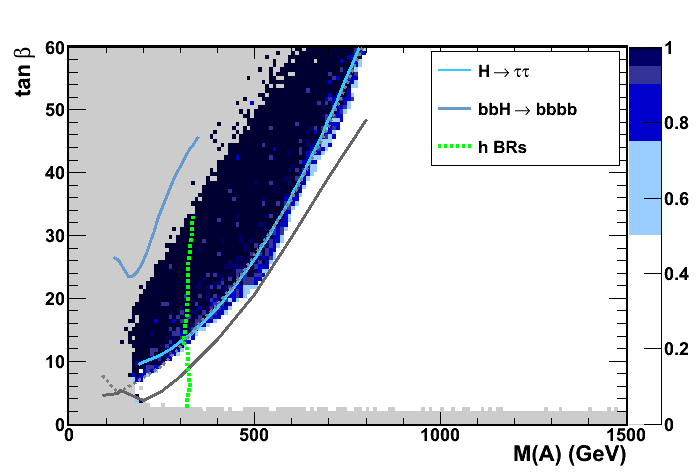} \\
\includegraphics[width=0.975\columnwidth]{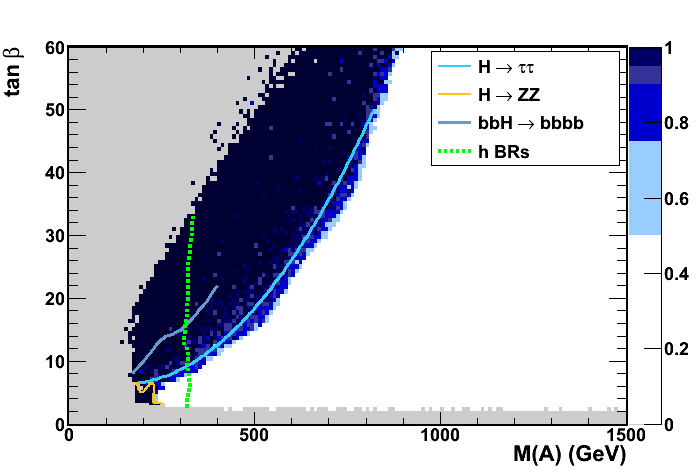} \\
\end{tabular}
\vspace*{-0.40cm}
\caption{Combination of the expected constraints on the $[M_A - \tan \beta]$ parameter plane from the $\tau \tau$ and $ZZ$ 
channels for (a) the current results (upper panel) and (b) their extrapolation to the full 8~TeV data set (lower panel). 
The colour scale gives the fraction of pMSSM points excluded at each $M_A$ and $\tan \beta$ value. The contours show the 
limits corresponding to 95\% or more of the points excluded. The 90\% C.L.\ constraint from the Higgs signal strengths is 
also shown. The expected and observed upper limits on $\tan \beta$ obtained in the MSSM $M_h^{max}$ scenario from the 
$\tau \tau$ channel search of~\cite{CMS-PAS-HIG-12-050} are indicated by the grey dotted and continuous lines, respectively, 
on the upper plot. The grey region has no accepted pMSSM points after the $B_s \rightarrow \mu \mu$, direct DM searches and 
$M_h$ constraints.}
\label{fig:Sum8TeV}
\end{figure}

First, we study the value of the product of production cross section and decay branching fraction in several channels 
by scanning over the pMSSM parameters. Figure~\ref{fig:XSBR8} shows the regions of the $[M_A - \tan \beta]$ parameter 
space where the product $\sigma \times \mathrm{BR}$ exceeds 1, 10 and 100~fb at 8~TeV for the 
$gg \to H/A$ and $bb \to H/A$ production processes and the $H/A \rightarrow ZZ$, $H/A \rightarrow WW$ and  
$H/A \rightarrow tt$ decays.
Finally, we combine the constraints derived in the various channels. We take the expected upper limits on the products 
$\sigma \times \mathrm{BR}$ in the various channels for both (a) the present status of the results and (b) their extrapolation to 
the full 8~TeV data set of 25~fb$^{-1}$. When limits are only available for the 7~TeV data set, we compute the expected limit at 
8~TeV by taking the ratio of production cross sections at the two energies, as a function of the $H/A$ mass, into account. 
For each channel, we consider the contours in the $[M_A - \tan \beta]$ plane where more than 95\% of the selected pMSSM points 
are excluded by these constraints. Alongside the $\tau \tau$ channel, the $ZZ$ and $bbbb$ channels also offer 
sensitivity on the 7 and 8~TeV data. 
For the $ZZ$ channel we use the upper limits on $\sigma \times \mathrm{BR}$ from~\cite{CMS-PAS-HIG-12-041}. These limits define 
an excluded region which connects with the $\tau \tau$ constraint at low masses and extends up to $M_A \simeq$ 550~GeV for 
$\tan \beta=$3-4. The $bbbb$ channels is based on the preliminary result on the 7~TeV CMS data~\cite{Behr:2013ji} extrapolated 
to the full 8~TeV data set.
We include also the constraint derived from the signal strengths, $\mu$, obtained in the ATLAS and CMS SM Higgs analyses for 
the $\gamma \gamma$, $WW$, $ZZ$ channels and the limits for $bb$ and $\tau \tau$, as discussed above, interpreting the observed 
particle as the SUSY lightest Higgs, $h$.
Results are summarised in Figure~\ref{fig:Sum8TeV}. The combination of the $H/A \rightarrow \tau \tau$
channel and the mass and $\mu$ values for the lightest $h$ boson exclude the region with $M_A >$ 320~GeV for all values of $\tan \beta$.
For the current results, the $\mu$ values defines this lower bound in the region of $\tan \beta$ = 2-15, where the direct search 
sensitivity is weaker. The sensitivity of the direct $H/A$ searches should approach this bound down to $\tan \beta \simeq$ 10, 
once the full 2012 data is analysed. The $ZZ$ channel, and to a lesser extent the $WW$, should close the low $M_A$ corner from 
$\tan \beta \simeq$ 2 up to the $\tau \tau$ limit for $M_A \lsim$ 230~GeV with the full 8~TeV data. The upper limits from the 
$t \bar t$ channel, for which only results at 7~TeV have been reported, are still below the expected values for $H/A$ production 
in the MSSM, even by extrapolating them to the full 8~TeV data set. Instead, this channel will become essential at 
13 and 14~TeV. The combination of these constraints from the Higgs sector provide limits on $M_A$ and $\tan \beta$, 
which are significantly  tighter compared to those derived from flavour physics, such as the BR($B_s \rightarrow \mu \mu$) for 
which the first measurement has recently been reported by LHCb~\cite{Aaij:2012nna} (see Figure~\ref{fig:Sum8TeV}).

\subsection{Perspectives at 14~TeV}

\begin{figure*}[ht!]
\vspace*{-0.20cm}
\begin{tabular}{ccc}
\hspace*{-0.25cm} \includegraphics[width=0.75\columnwidth]{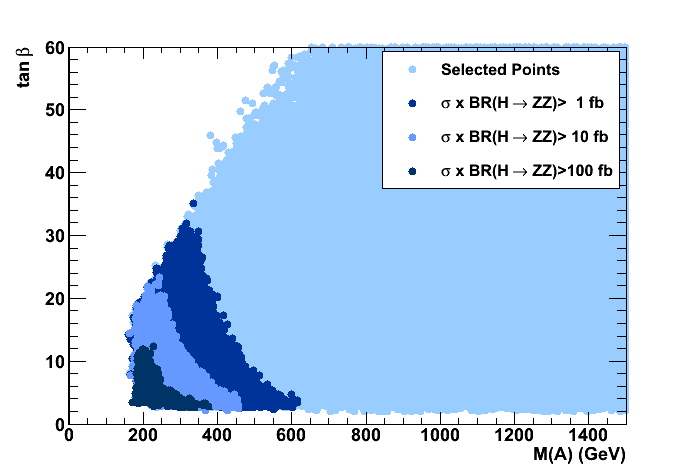} &
\hspace*{-0.75cm} \includegraphics[width=0.75\columnwidth]{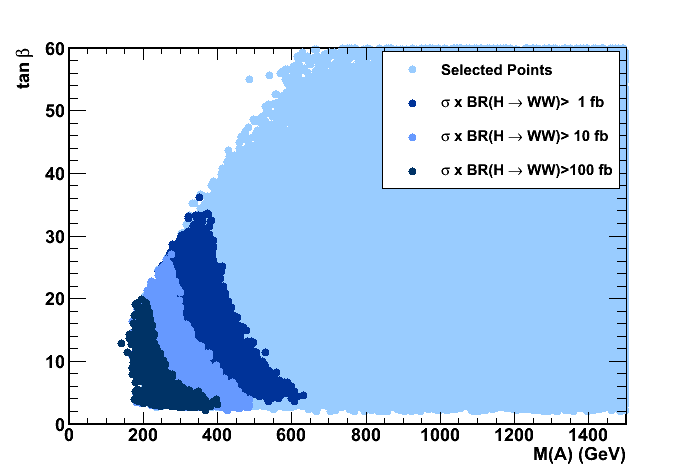} &
\hspace*{-0.75cm} \includegraphics[width=0.75\columnwidth]{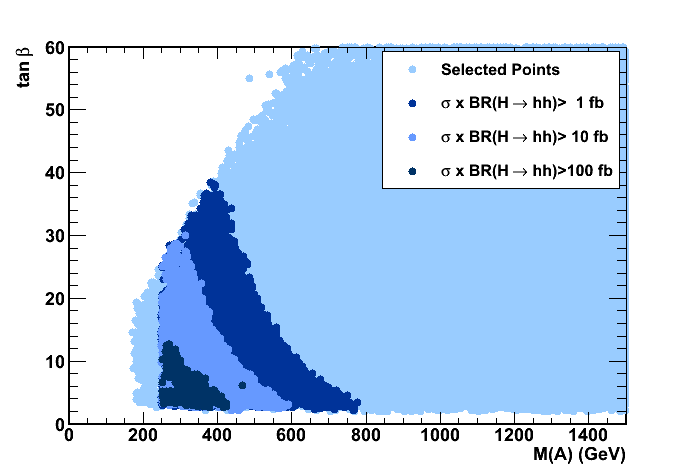}\\
\hspace*{-0.25cm} \includegraphics[width=0.75\columnwidth]{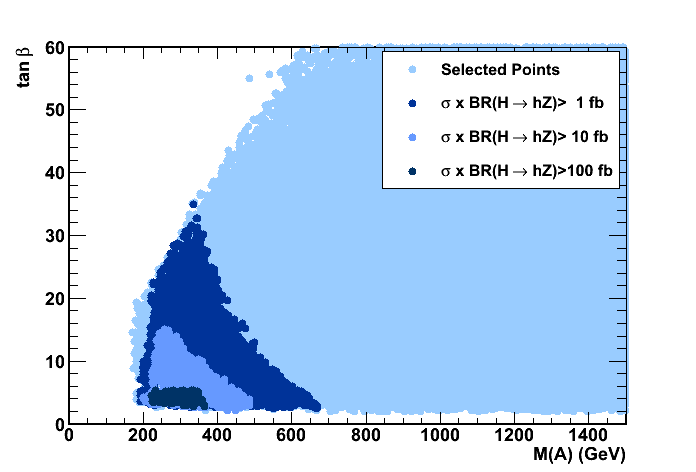} &
\hspace*{-0.75cm} \includegraphics[width=0.75\columnwidth]{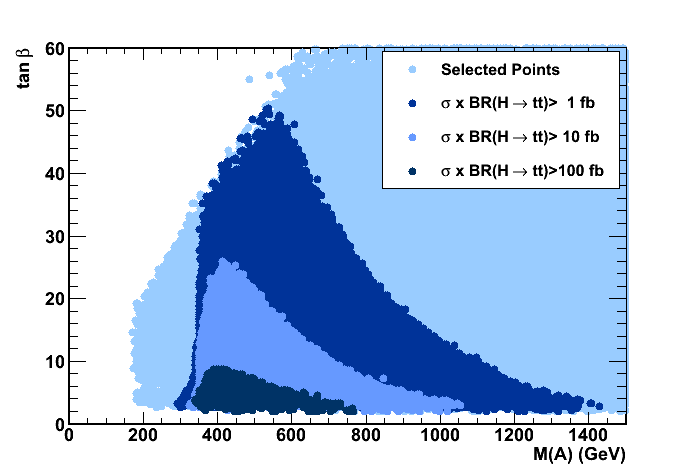} &
\hspace*{-0.75cm} \includegraphics[width=0.75\columnwidth]{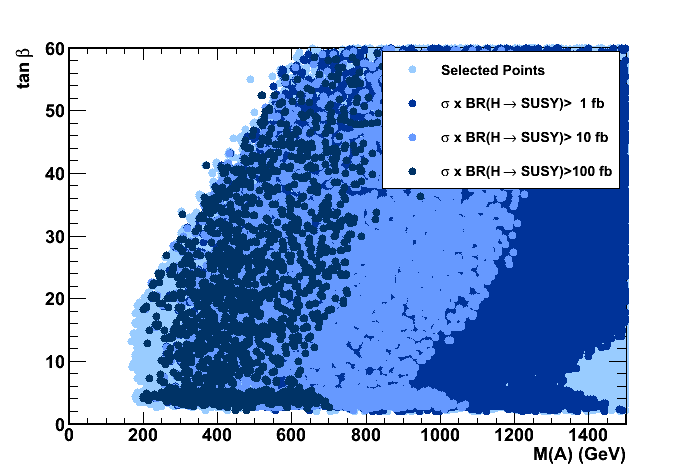} \\
\end{tabular}
\vspace*{-0.40cm}
\caption{Product of production cross section and decay branching fraction for $H \rightarrow ZZ$ (upper left), 
$H \rightarrow WW$ (upper centre), $H \rightarrow hh$ (upper left), $A \rightarrow hZ$ (bottom right), 
$H \rightarrow t \bar{t}$ (bottom centre) and $H \rightarrow$ SUSY particles (bottom left),  
at 14~TeV in the $[M_A - \tan \beta]$ parameter plane. The colour coding is given in the legend and it 
is the same as in Figure~\ref{fig:XSBR8}.}
\label{fig:XSBR14}
\end{figure*}

\begin{figure}[hb!]
\vspace*{-0.35cm}
\includegraphics[width=0.975\columnwidth]{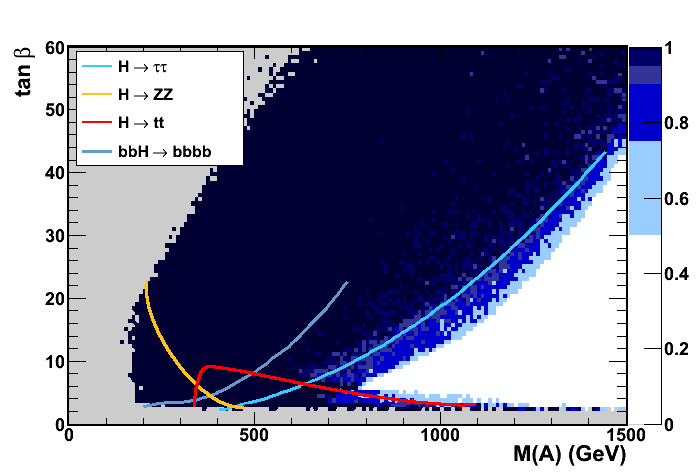}
\vspace*{-0.40cm} 
\caption{Combination of the expected constraints on the $[M_A - \tan \beta]$ parameter plane from the $\tau \tau$, $ZZ$ 
and $tt$ channels as in Figure~\ref{fig:Sum8TeV}, extrapolated to 150~fb$^{-1}$ at 14~TeV. The colour scale gives the 
fraction of pMSSM points excluded at each $M_A$ and $\tan \beta$ value. The grey region has no accepted pMSSM 
points after the $B_s \rightarrow \mu \mu$, direct DM searches and $M_h$ constraints.}
\label{fig:Sum14TeV}
\end{figure}

The increase of the production cross sections moving from 7 to 14~TeV is a factor of 4.5 to 9 for $gg \rightarrow H/A$ 
and 5 to 12 for $bb \rightarrow H/A$ in the mass range 300 to 800~GeV. 
Figure~\ref{fig:XSBR14} shows the regions of the $[M_A - \tan \beta]$ parameter space where the product 
$\sigma \times \mathrm{BR}$ exceeds 1, 10 and 100~fb for the $gg \to H/A$ and $bb \to H/A$ production processes 
and the $H/A \rightarrow ZZ$, $H/A \rightarrow WW$, $H/A \rightarrow hh$, $A \rightarrow hZ$, 
 $H/A \rightarrow t \bar t$ and the inclusive decays $H/A \rightarrow$ SUSY particles.
At the high mass end the product 
$\sigma \times \mathrm{BR}$ of $\sim$10~fb, corresponding to the current sensitivity at 800~GeV in the $\tau \tau$ channel, 
is obtained beyond $M_A=$ 1~TeV. At 13 and 14~TeV the sensitivity extends to mass values above the $hh$, $hZ$ and the $tt$ 
production thresholds at small to intermediate values of $\tan \beta$, which make these channels 
relevant to the LHC searches. In this region the $\tau \tau$ 
channel alone cannot ensure the coverage of the $[M_A - \tan \beta]$ plane and these additional channels need to be included. 
The $ZZ$ channel provides redundancy while the $tt$ decay is most important, in particular at large $M_A$ and low $\tan \beta$ 
values. The $WW$ channel has more limited interest, since its sensitivity is lower than $ZZ$. The combination of the $\tau \tau$, 
$ZZ$ and $tt$ modes covers the $[M_A - \tan \beta]$ parameter plane up to $M_A \simeq$ 700~GeV for any value of $\tan \beta$, 
as shown in Figure~\ref{fig:Sum14TeV}. 

\subsubsection{Characterisation of $hZ$ and $hh$ channels}

The decays $H \rightarrow hh$ and $A \rightarrow hZ$ are important in providing redundancy at low values of $\tan \beta$ and 
intermediate $M_A$ masses. 
They also result in rather distinctive $bbbb$, $bb \tau \tau$ and $bb \ell \ell$ ($\ell = e$, $\mu$) final states, which should 
be investigated in the high energy LHC runs. Since these modes have not yet been searched for in the LHC data, we characterise 
here their decay kinematics and study the reconstruction strategies using a simple analysis for signal events. 

\begin{figure}[h!]
\vspace*{-0.20cm}
\begin{tabular}{cc}
\hspace*{-0.30cm} \includegraphics[width=0.55\columnwidth]{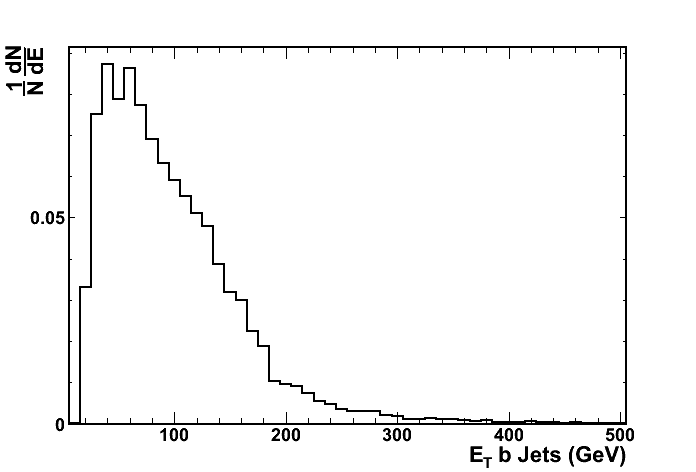} &
\hspace*{-0.60cm} \includegraphics[width=0.55\columnwidth]{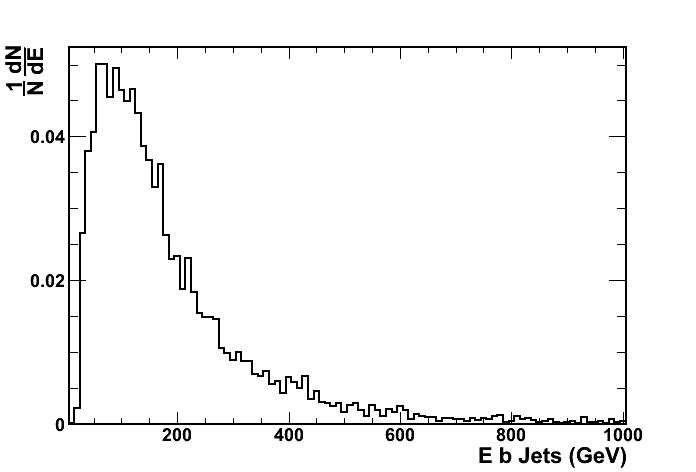} \\
\vspace*{-0.25cm} & \vspace*{-0.25cm} \\
\hspace*{-0.30cm} \includegraphics[width=0.55\columnwidth]{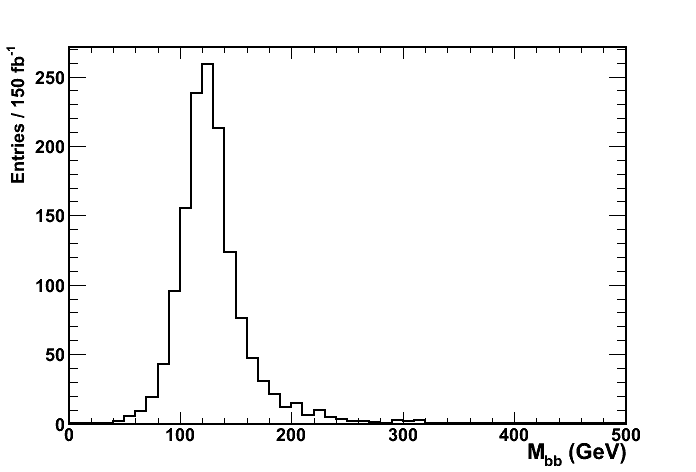} &
\hspace*{-0.60cm} \includegraphics[width=0.55\columnwidth]{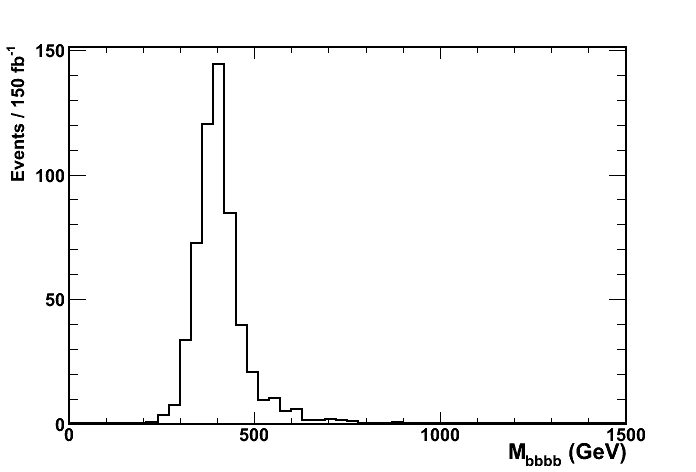} \\
\end{tabular}
\vspace*{-0.40cm}
\caption{Reconstruction of  $H \rightarrow hh \rightarrow bbbb$ events at 14~TeV for $M_H$ = 400~GeV: distribution of the 
$b$-jet transverse energy $E_T$ (upper right) and energy $E$ (lower left),  invariant mass of $bb$ pairs (lower left) and 
$bbbb$ invariant mass (lower right). A BR($H \rightarrow hh$) = 0.12 has been assumed.}
\label{fig:Hhh14TeV}
\end{figure}

\begin{figure}[h!]
\vspace*{-0.20cm}
\begin{tabular}{cc}
\hspace*{-0.30cm} \includegraphics[width=0.55\columnwidth]{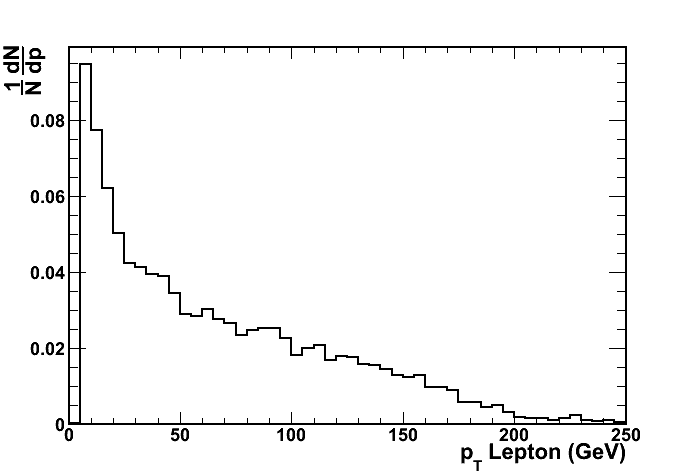} &
\hspace*{-0.60cm} \includegraphics[width=0.55\columnwidth]{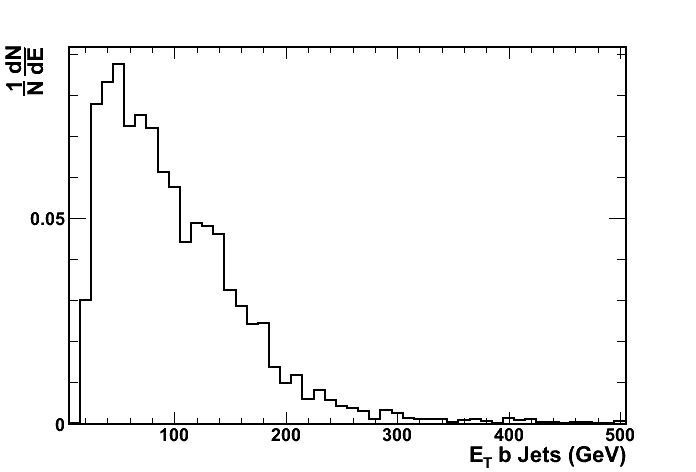} \\
\vspace*{-0.25cm} & \vspace*{-0.25cm} \\
\hspace*{-0.30cm} \includegraphics[width=0.55\columnwidth]{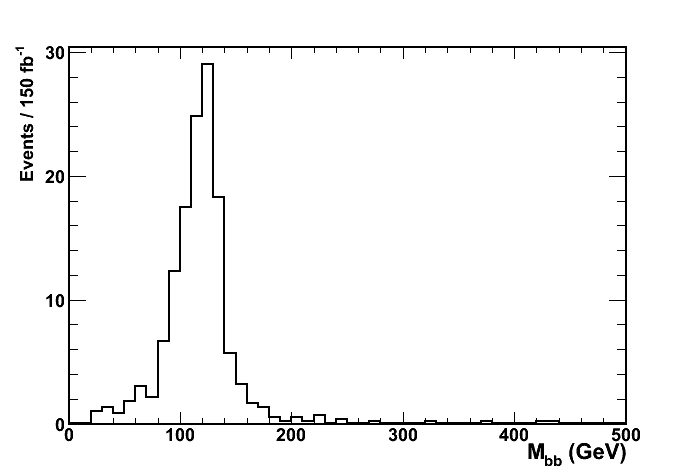} &
\hspace*{-0.60cm} \includegraphics[width=0.55\columnwidth]{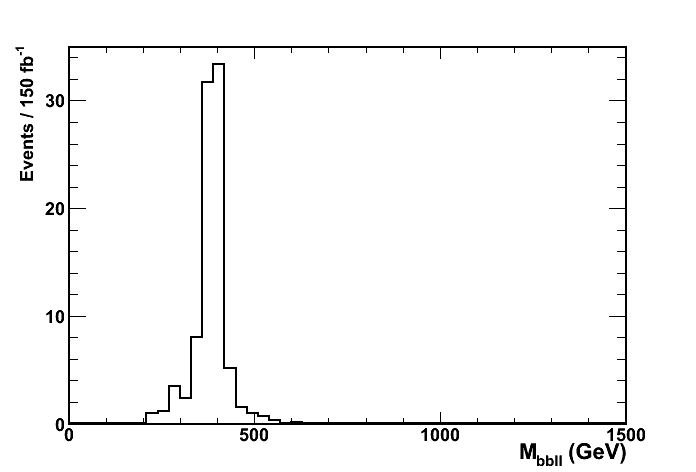} \\
\end{tabular}
\vspace*{-0.40cm}
\caption{Reconstruction of  $A \rightarrow Zh \rightarrow  \ell \ell bb$ events at 14~TeV for $M_A$ = 400~GeV: distribution of 
the lepton transverse energy   $p_{T}$ (upper left), $b$-jet transverse energy $E_T$ (upper right), $bb$ (lower left) and $bbll$ 
(lower right) invariant mass. A BR($A \rightarrow Zh$) = 0.12 has been assumed.}
\label{fig:AZh14TeV}
\end{figure}

These events are generated using {\tt Pythia 8.1}~\cite{pythia8} at 14~TeV and scaled to an integrated luminosity of 150~fb$^{-1}$. 
For this study, we have chosen $M_A$ = 400 and 500~GeV, $\tan \beta=$ 5 with branching fractions of 0.12 for $H \rightarrow hh$ 
and $A \rightarrow hZ$.
The detector response simulation is performed using {\tt Delphes 3.0}~\cite{Ovyn:2009tx}. Jets are reconstructed using the anti-kt 
algorithm~\cite{Cacciari:2008gp} implemented in {\tt FastJet}~\cite{fastjet}, requiring their pseudorapidity, $\eta$, not to exceed 
2.8 and transverse momentum $p_T>$20~GeV. 
Electrons and muons are accepted for $|\eta|<$2.4 and $p_T>$20~GeV. $b$-jets are accepted at $\eta <$ 2.5, assuming a tagging 
efficiency of 75\% per jet. In both channels, $b$ jets are rather soft, with the transverse energy distributions peaking around 
50~GeV, thus emphasising $b$ tagging at relatively small transverse energies (see Figures \ref{fig:Hhh14TeV} and \ref{fig:AZh14TeV}). 
Similarly low is the transverse energy distribution of leptons from the $Z$ decay in the $A$ channel, which has its most probable 
value just above the $p_T$ cut applied in this analysis (see Figure~\ref{fig:AZh14TeV}).

$H \rightarrow hh \rightarrow bbbb$ events are reconstructed by requiring at least three $b$-tagged jets. 
The pairing of four $b$ jets, or three $b$ jets with any of the reconstructed jets, which minimises the mass difference of the 
two di-jet pairs and their difference from the $h$ mass of 126~GeV is selected. The di-jet invariant mass distribution is shown 
in Figure~\ref{fig:Hhh14TeV}. The invariant mass resolution obtained with the fast simulation is comparable to that reported for 
the $H_{\mathrm{SM}} \rightarrow b \bar b$ search. The four-jet invariant mass, $M_{bbbb}$ shows a clear peak corresponding 
to the generated $H$ mass as shown in Figure~\ref{fig:Hhh14TeV}. The efficiency of this selection for the signal mass region of 
300$< M_{bbbb} <$ 500~GeV is $\simeq$ 16\% at both values of $M_H$.

For the $Zh \rightarrow \ell \ell bb$ we select events with two, oppositely charged, electrons or muons with two or 
more jets, of which at least one $b$ tagged. The $\ell \ell$ invariant mass is required to be consistent with that of the $Z$ 
within the resolution. 
If the event contains exactly two $b$-tagged jets, the invariant mass of the pair is required 
to be consistent with 126~GeV within the resolution. If there is only one $b$-tagged jet, but it has a mass consistent with 126~GeV, 
this is also accepted. The final mass is computed by combining the di-leptons with the di-jet pair or the single $b$ jet. The 
resulting distribution is shown in Figure~\ref{fig:AZh14TeV} for an integrated luminosity of 150~fb$^{-1}$. The selection 
efficiency for the loose signal mass region of 300$< M_{bbll} <$ 500~GeV is $\simeq$ 25\% at both values of $M_A$.

\begin{figure}[hb!]
\vspace*{-0.25cm}
\includegraphics[width=0.975\columnwidth]{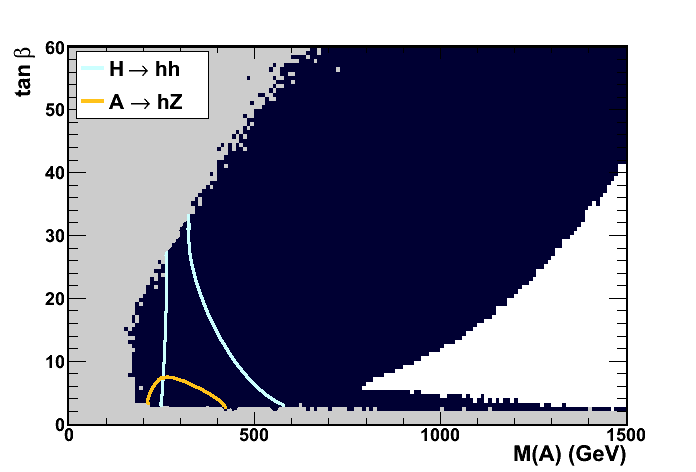}
\vspace*{-0.40cm} 
\caption{Regions on the $[M_A - \tan \beta]$ parameter plane where the $H \rightarrow hh$ and $A \rightarrow hZ$ process 
yield 50 reconstructed events for 150~fb$^{-1}$ at 14~TeV, compared with the coverage provided by the combination of 
$\tau \tau$ and $ZZ$ shown in dark blue. The grey region has no accepted pMSSM points after the $B_s \rightarrow \mu \mu$, 
direct DM searches and $M_h$ constraints.}
\label{fig:SumAddTeV}
\end{figure}
Since we base this preliminary characterisation on the reconstruction of signal only event and have not considered the backgrounds, 
we cannot define here exclusion contours. Instead, we simply plot the regions of the $[M_A - \tan \beta]$ plane where we register more 
than 50 reconstructed events for 150~fb$^{-1}$ of data at 14~TeV. The result is shown in Figure~\ref{fig:SumAddTeV}, where the 
region covered by the $hh$ and $hZ$ final states is compared to that of expected sensitivity for the combination of the $\tau \tau$, 
$ZZ$ and $tt$ channels, considered above. We notice that the $hh$ channel covers the full $\tan \beta$ range of interest from 
threshold up to $M_A \simeq$ 400~GeV and up to 550~GeV at low $\tan \beta$ values, beyond the $ZZ$ sensitivity. In this important region of 
small to intermediate values of $\tan \beta$, the $hh$ and $hZ$ channels provide redundancy to the coverage offered by the $\tau \tau$ 
and $t \bar t$ modes.

\subsection{Effect of QCD Uncertainties and SUSY Particles}

The limits derived above do not account for the effects of theoretical uncertainties, affecting the Higgs production 
cross section and decay branching fractions, and of SUSY contributions.  First, the $gg \to H/A$ and $b\bar b \to H/A$ 
cross sections have sizeable QCD uncertainties from the factorisation and renormalisation scales, parton distribution 
functions (PDFs) and parametric systematics from $\alpha_s$ and the heavy quark masses. 
We estimate the parametric systematics on the cross section for $\alpha_s = 0.118\pm0.0012$, $\bar{m_b} (\bar{m_b})$ = 
(4.19$\pm$0.05)~GeV, $m_t$ = (172.9$\pm$1.5)~GeV and those from the PDFs by taking the largest difference between 
different sets of functions. The latter is the dominant contribution.
%and follow the LHC Higgs cross section working group prescriptions for the effect of 
%the PDFs~\cite{LHCHiggsCrossSectionWorkingGroup:2011ti}. 
The combination of the uncertainties on the quark masses, PDFs and $\alpha_s$ leads to an estimated systematic 
uncertainty on the $pp \to H/A$ rate of $\approx \pm 24\%$ at 8~TeV and $\approx \pm 20\%$ at 14~TeV, dominated by 
the PDFs and scale, and comparable to those for 
$pp \rightarrow H_{\mathrm{SM}}$ production~\cite{LHCHiggsCrossSectionWorkingGroup:2011ti,Dittmaier:2012vm}. 

\begin{figure}[ht!]
\begin{center}
\vspace*{-0.20cm}
\begin{tabular}{cc}
\hspace*{-0.30cm} \includegraphics[width=0.55\columnwidth]{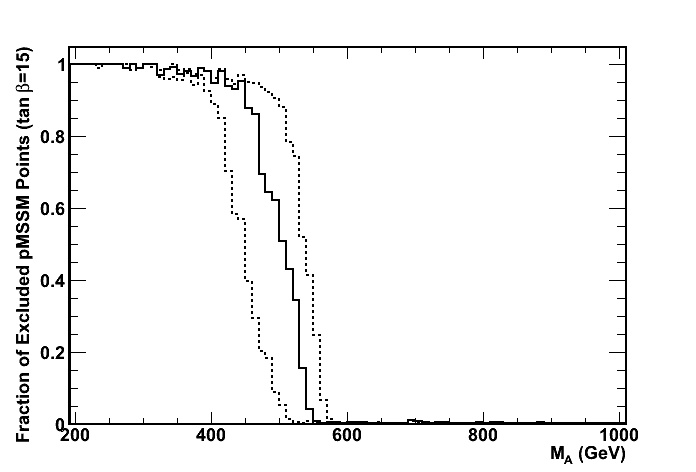} &
\hspace*{-0.70cm} \includegraphics[width=0.55\columnwidth]{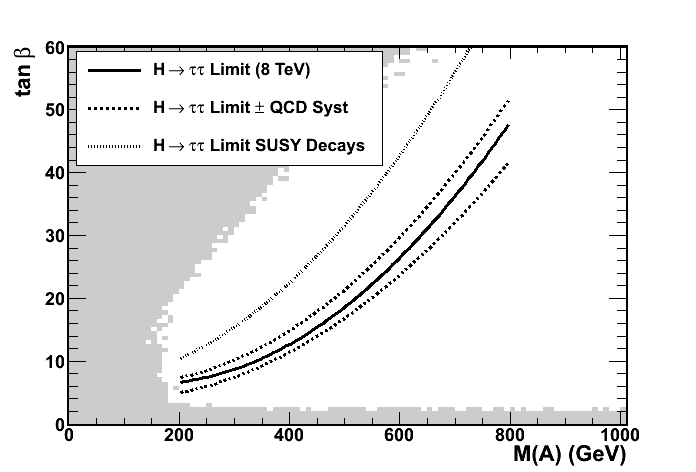} \\
\end{tabular}
\end{center}
\vspace*{-0.40cm}
\caption{QCD systematics and SUSY particle effects on the projected $H/A \rightarrow \tau \tau$ exclusion with 
at 8~TeV. Left: fraction of pMSSM points excluded for $\tan \beta$ = 15 as a function 
of $M_A$ (continuous line) and the effect of a change by $\pm$25\% of the $pp \rightarrow H/A$ production cross section 
to reflect QCD uncertainties (dashed lines). Right: limits from the $\tau \tau$ in the $[M_A - \tan \beta]$ plane obtained 
by varying the production cross section (dashed lines) and requiring less than 0.1\% of the points around the limit to fail 
exclusion due to the effect of SUSY decays. The grey region has no accepted pMSSM points after the $B_s \rightarrow \mu \mu$, 
direct DM searches and $M_h$ constraints.}
\label{fig:H0tau08}
\end{figure}

\begin{figure}[ht!]
\begin{center}
\vspace*{-0.20cm}
\begin{tabular}{cc}
\hspace*{-0.30cm} \includegraphics[width=0.55\columnwidth]{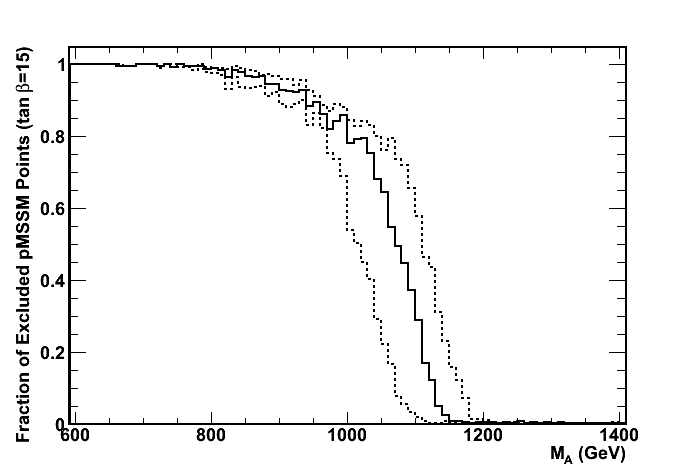} &
\hspace*{-0.70cm} \includegraphics[width=0.55\columnwidth]{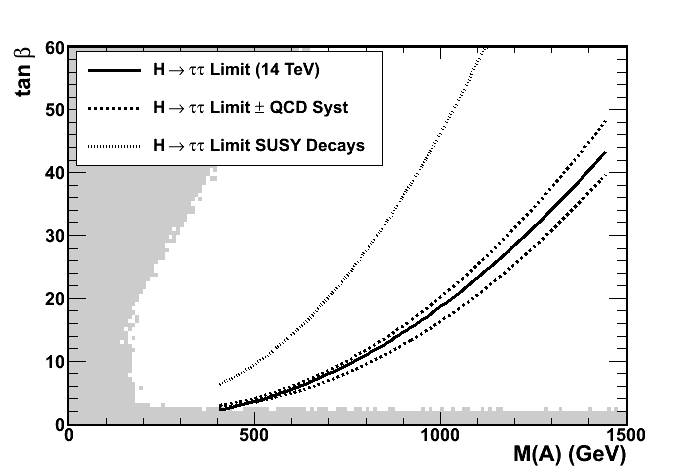} \\
\end{tabular}
\end{center}
\vspace*{-0.40cm}
\caption{QCD systematics and SUSY particle effects on the projected $H/A \rightarrow \tau \tau$ exclusion, as in 
Figure~\ref{fig:H0tau08}, for 150~fb$^{-1}$ at 14~TeV.}
\label{fig:H0tau14}
\end{figure}

In order to evaluate their impact on the exclusion contours in the $[M_A - \tan \beta]$ plane, we repeat 
our study while changing the production cross section by $\pm$25\% and compare the constraints obtained to that
corresponding to the central values for the production cross sections. Figures~\ref{fig:H0tau08} and \ref{fig:H0tau14} 
show the fractions of excluded points in the $[M_A - \tan \beta]$ plane and their projections as a function of $M_A$ for the 
fixed value of $\tan \beta$ = 15 at 8 and 14~TeV, respectively, and includes the effect of the $\pm \sigma_{\mathrm{QCD}}$ 
change of the cross sections by the QCD uncertainties. The effect is a shift of the excluded $M_A$ mass by 
$\pm$45~GeV at 8~TeV and by $\pm$55~GeV at 14~TeV at $\tan \beta$=15 and larger for higher values of $\tan \beta$.

Then, we observe that, there is a significant smearing of the curve giving the fraction of excluded pMSSM points as a 
function of $M_A$, even if the systematics on the production cross section are ignored. In fact,  
the exclusion curve goes from 10\% to 90\% of the points excluded over a range of $M_A$ values spanning $\sim$90~GeV 
at 8~TeV and $\sim$150~GeV at 14~TeV, as a result of the variation of other pMSSM parameters. This range, 
which is comparable to that corresponding to the QCD uncertainty obtained above, is intrinsic to the pMSSM and includes 
contributions such as the loop effect through the $\Delta_b$ term discussed in section II.B. 

Finally, we consider quantitatively the region of the $[M_A - \tan \beta]$ plane where decays into SUSY channels may 
invalidate the $\tau \tau$ limit. The panels on the right in Figures \ref{fig:H0tau08} and \ref{fig:H0tau14}, have dashed 
lines showing the limit of the region where this may occur, for 8 and 14~TeV, respectively. Since the 
$\sigma \times$BR($H/A \rightarrow \tau \tau$) product increases for low $M_A$ and high $\tan \beta$ values, there is a 
region where the SUSY decays cannot upset the exclusion obtained in this channel, since the $H \rightarrow$ SUSY branching 
fraction is $\lsim$0.60, as shown in Figure~\ref{fig:BRHSUSY}. However, the region 
affected by the SUSY decays extends much further towards lower $A$ masses compared to that describing the effect of the QCD 
and parametric uncertainties on the production cross section and also the SUSY loop effects.
The width of this region, $\sim$150~GeV at 8~TeV for $\tan \beta=$15, and the occurrence of these points increase with the 
energy which gives access to heavier bosons with decays into pairs of SUSY particles kinematically allowed. 
Moving from 8 to 14~TeV, the width of the regions doubles and the occurrence of these points increase by a factor of $\sim$1.5. 
%From our scans, we register a 1\% of pMSSM points with $M_A$ and $\tan \beta$ values which are excluded in the MSSM 
%$M_h^{max}$ model used by ATLAS and CMS for setting their 95\% C.L. upper limits on $\tan \beta$, but correspond to 
%values of product $\sigma \times \mathrm{BR(}H/A \rightarrow \tau \tau \mathrm{)}$ still below the CMS expected limit of 
%Ref.~\cite{CMS-PAS-HIG-12-050} and thus are not excluded. 
The occurrence of the various SUSY effects we have discussed for the pMSSM points within the region of the  $[M_A - \tan \beta]$ plane 
excluded in the MSSM $M_h^{max}$ model but not excluded due to their low value of the $\sigma \times$BR($H/A \rightarrow \tau \tau$) 
product has been studied for our scans. We observe that $H/A$ decays into $\tilde{\chi} \tilde{\chi}$ pairs are responsible for 
55\% of the cases and those into $\tilde{\tau} \tilde{\tau}$ pairs for another 10\%, while it is a $\Delta_b$ term large and negative 
in sign which suppresses the $\tau \tau$ rate for the remaining 35\% of the points failing exclusion. 
When the $H$ decays into $\tilde{\chi} \tilde{\chi}$ pair, the dominant state is the lightest neutralino $\tilde{\chi^0_1}$ in about 
1/4 of the cases. In the other cases, we observe an increase in the yield of $\tilde{\chi}^0_{2,3} \rightarrow h/Z \tilde{\chi}^0_1$, 
which may offer an important signature for the LHC searches~\cite{Arbey:2012fa}.

\section{Conclusions}

The search for heavy Higgs bosons represents a next frontier in the understanding of the Higgs sector after 
the discovery of the Higgs-like state at 126~GeV at the LHC and the first results on its decays, spin and parity.
The combination of the indirect limits from the $h$ signal strengths and the direct searches in the $\tau \tau$ 
and $ZZ$ channels should impose an exclusion limits in $[M_A - \tan \beta]$ plane around $M_A \gsim$ 320~GeV for 
the 7+8~TeV data, determined by the indirect limit from the rates of the observed Higgs boson

As the mass sensitivity of the LHC searches increases with the energy and integrated luminosity, more final states 
than $\tau \tau$ become relevant to effectively constrain the supersymmetric parameter space, in particular at low to
moderate values of $\tan \beta$. In fact, low values of $\tan \beta$ are still viable, after incorporating the $M_h$ 
constraint, provided high SUSY scales, $M_S$, are chosen and they represent a scenario, rich in decays into $t \bar t$ 
and $ZZ$, $hh$ and $hZ$ boson pairs, which should be carefully explored at the LHC at 13 and 14~TeV. 
The $M_S$ bound will reach $M_A \gsim$ 800~GeV for any value of $\tan \beta$ with 150~fb$^{-1}$ of data at 14~TeV, 
determined by the direct searches for heavy Higgs states.
The effects of the SUSY particle spectrum, other SUSY parameters and the QCD theoretical uncertainties need to be carefully 
considered. SUSY loops and QCD effects on the $M_A$ bounds are find to be quite comparable in size. However, scenarios where 
decays into SUSY particles are important, or even dominant, exist and these channels need to be accounted for in the LHC 
searches at 13 and 14~TeV. 

The constraints derived by the study of the Higgs sector are becoming an essential part of the probe of the SUSY parameter 
space at the LHC and offer an essential complement to the searches for strongly interacting SUSY particles and gauginos.  

\begin{acknowledgments}
We are grateful to Abdelhak Djouadi who provided us with inspiration for several parts of this study, in particular 
the interest for the low $\tan \beta$ region and the use of the $ZZ$ and $tt$ analyses to constrain heavy Higgs 
production. We wish to thank also Benjamin Allanach for discussion and support with the use of Softsusy, 
Robert Harlander with that of SusHi and Stefan Dittmaier for discussion on the QCD systematics in Higgs production.
M.B. wishes to thank the Galileo Galilei Institute for Theoretical Physics for the hospitality and INFN for partial 
support during the early stages of this work.
\end{acknowledgments}


\begin{thebibliography}{99}

\bibitem{Aad:2012tfa}
   G.~Aad {\it et al.}  [ATLAS Collaboration],
   %``Observation of a new particle in the search for the Standard Model
   %Higgs boson with the ATLAS detector at the LHC,''
   Phys.\ Lett.\ B {\bf 716} (2012) 1
   [arXiv:1207.7214 [hep-ex]].

\bibitem{Chatrchyan:2012ufa}
   S.~Chatrchyan {\it et al.}  [CMS Collaboration],
   %``Observation of a new boson at a mass of 125 GeV with the CMS
   %experiment at the LHC,''
   Phys.\ Lett.\ B {\bf 716} (2012) 30
   [arXiv:1207.7235 [hep-ex]].

\bibitem{Baglio:2011xz}
  J.~Baglio and A.~Djouadi,
  %``Implications of the ATLAS and CMS searches in the channel $pp \to Higgs \to \tau^+\tau^$ for the MSSM and SM Higgs bosons,''
  arXiv:1103.6247 [hep-ph].
  %%CITATION = ARXIV:1103.6247;%%

\bibitem{Carena:2011fc}
  M.~Carena, P.~Draper, T.~Liu and C.~Wagner,
  %``The 7 TeV LHC Reach for MSSM Higgs Bosons,''
  Phys.\ Rev.\ D {\bf 84} (2011) 095010
  [arXiv:1107.4354 [hep-ph]].
  %%CITATION = ARXIV:1107.4354;%%

\bibitem{Christensen:2012ei}
  N.~D.~Christensen, T.~Han and S.~Su,
  %``MSSM Higgs Bosons at The LHC,''
  Phys.\ Rev.\ D {\bf 85} (2012) 115018
  [arXiv:1203.3207 [hep-ph]].
  %%CITATION = ARXIV:1203.3207;%%

\bibitem{Chang:2012zf}
  J.~Chang, K.~Cheung, P.~-Y.~Tseng and T.~-C.~Yuan,
  %``Implications on the Heavy CP-even Higgs Boson from Current Higgs Data,''
  Phys.\ Rev.\ D {\bf 87} (2013) 035008
  [arXiv:1211.3849 [hep-ph]].
  %%CITATION = ARXIV:1211.3849;%%

\bibitem{Carena:2013qia}
  M.~Carena, S.~Heinemeyer, O.~Stal, C.~E.~M.~Wagner and G.~Weiglein,
  %``MSSM Higgs Boson Searches at the LHC: Benchmark Scenarios after the Discovery of a Higgs-like Particle,''
  arXiv:1302.7033 [hep-ph].
  %%CITATION = ARXIV:1302.7033;%%

\bibitem{Djouadi:1998di}
  A.~Djouadi {\it et al.}  [Les Houches MSSM Working Group],
  %``The Minimal supersymmetric standard model: Group summary report,''
  hep-ph/9901246.
  %%CITATION = HEP-PH/9901246;%%

\bibitem{ATLASTDR}
  [ATLAS Collaboration], CERN/LHCC 99-14.

\bibitem{Djouadi:2005gj}
  A.~Djouadi,
  %``The Anatomy of electro-weak symmetry breaking. II. The Higgs bosons in the minimal supersymmetric model,''
  Phys.\ Rept.\  {\bf 459} (2008) 1
  [hep-ph/0503173].

\bibitem{Muhlleitner:2010zz}
  M.~Muhlleitner, H.~Rzehak and M.~Spira,
  %``MSSM Higgs boson production via gluon fusion,''
  DESY-PROC-2010-01.
  %%CITATION = DESY-PROC-2010-01;%%

\bibitem{Arbey:2012dq}
  A.~Arbey, M.~Battaglia, A.~Djouadi and F.~Mahmoudi,
  %``The Higgs sector of the phenomenological MSSM in the light of the Higgs boson discovery,''
  JHEP {\bf 1209} (2012) 107 [arXiv:1207.1348 [hep-ph]].
  %%CITATION = ARXIV:1207.1348;%%

\bibitem{Arbey:2012bp}
  A.~Arbey, M.~Battaglia, A.~Djouadi and F.~Mahmoudi,
  %``An update on the constraints on the phenomenological MSSM from the new LHC Higgs results,''
  Phys.\ Lett.\ B {\bf 720} (2013) 153
  [arXiv:1211.4004 [hep-ph]].
  %%CITATION = ARXIV:1211.4004;%%

\bibitem{Arbey:2011aa}
  A.~Arbey, M.~Battaglia and F.~Mahmoudi,
  %``Constraints on the MSSM from the Higgs Sector: A pMSSM Study of Higgs Searches, $B^0_s -> \mu^+ \mu^-$ and Dark Matter Direct Detection,''
  Eur.\ Phys.\ J.\ C {\bf 72} (2012) 1906
  [arXiv:1112.3032 [hep-ph]].
  %%CITATION = ARXIV:1112.3032;%%

\bibitem{Arbey:2012ax}
  A.~Arbey, M.~Battaglia, F.~Mahmoudi and D.~Martinez Santos,
  %``Supersymmetry confronts Bs -> mu+mu-: Present and future status,''
  Phys.\ Rev.\ D {\bf 87} (2013) 035026
  [arXiv:1212.4887 [hep-ph]].
  %%CITATION = ARXIV:1212.4887;%%

\bibitem{Aaij:2012nna}
  R.~Aaij {\it et al.}  [LHCb Collaboration],
  %``First evidence for the decay Bs -> mu+ mu-,''
  Phys.\ Rev.\ Lett.\  {\bf 110} (2013) 021801
  [arXiv:1211.2674 [hep-ex]].
  %%CITATION = ARXIV:1211.2674;%%

\bibitem{Aprile:2012nq}
  E.~Aprile {\it et al.}  [XENON100 Collaboration],
  Phys.\ Rev.\ Lett.\ {\bf 109} (2012) 181301.
%  arXiv:1207.5988 [astro-ph.CO].

\bibitem{Arbey:2011un}
  A.~Arbey, M.~Battaglia and F.~Mahmoudi,
  %``Implications of LHC Searches on SUSY Particle Spectra: The pMSSM Parameter Space with Neutralino Dark Matter,''
  Eur.\ Phys.\ J.\ C {\bf 72} (2012) 1847
  [arXiv:1110.3726 [hep-ph]].
  %%CITATION = ARXIV:1110.3726;%%

\bibitem{Djouadi:1997yw}
  A.~Djouadi, J.~Kalinowski and M.~Spira,
  Comput.\ Phys.\ Commun.\  {\bf 108} (1998) 56.

\bibitem{Spira:1995rr}
  M.~Spira, A.~Djouadi, D.~Graudenz and P.~M.~Zerwas,
  %``Higgs boson production at the LHC,''
  Nucl.\ Phys.\ B {\bf 453} (1995) 17
  [hep-ph/9504378].
  %%CITATION = HEP-PH/9504378;%%

\bibitem{Spira:1996if}
  M.~Spira,
  %``HIGLU and HDECAY: Programs for Higgs boson production at the LHC and Higgs boson decay widths,''
  Nucl.\ Instrum.\ Meth.\ A {\bf 389} (1997) 357
  [hep-ph/9610350].
  %%CITATION = HEP-PH/9610350;%%

\bibitem{bbh}
  R.V. Harlander and W.B. Kilgore,
  %''Higgs boson production in bottom quark fusion at next-to-next-to-leading order''
  Phys.\ Rev.\ D {\bf 68} (2003) 013001 [hep-ph/0304035].

\bibitem{Spira:1997dg}
   M.~Spira,
   %``QCD effects in Higgs physics,''
   Fortsch.\ Phys.\  {\bf 46} (1998) 203
   [hep-ph/9705337].

\bibitem{Harlander:2012pb}
  R.~V.~Harlander, S.~Liebler and H.~Mantler,
  %``SusHi: A program for the calculation of Higgs production in gluon fusion and bottom-quark annihilation in the Standard Model and the MSSM,''
  arXiv:1212.3249 [hep-ph].

\bibitem{softsusy}  
  B.C. Allanach, Comput. Phys. Commun. {\bf 143} (2002) 305
  [arXiv:hep-ph/0104145 [hep-ph]].

\bibitem{superiso}
   F.~Mahmoudi,
   Comput.\ Phys.\ Commun.\  {\bf 178} (2008) 745 [arXiv:0710.2067 [hep-ph]];
   {\it idem}, Comput.\ Phys.\ Commun.\  {\bf 180} (2009) 1579 [arXiv:0808.3144 [hep-ph]].

\bibitem{superiso_relic}
   A.~Arbey and F.~Mahmoudi,
   Comput.\ Phys.\ Commun.\  {\bf 181} (2010) 1277 [arXiv:0906.0369
     [hep-ph]].

\bibitem{Pierce:1996zz}
  D.~M.~Pierce, J.~A.~Bagger, K.~T.~Matchev, R.~-j.~Zhang and ,
  %``Precision corrections in the minimal supersymmetric standard model,''
  Nucl.\ Phys.\ B {\bf 491} (1997) 3
  [hep-ph/9606211].
  %%CITATION = HEP-PH/9606211;%%

\bibitem{Hollik:2011xd}
  W.~Hollik and J.~-H.~Zhang,
  %``Radiative Corrections to $H^0\to WW/ZZ$ in the MSSM,''
  Phys.\ Rev.\ D {\bf 84} (2011) 055022
  [arXiv:1109.4781 [hep-ph]].
  %%CITATION = ARXIV:1109.4781;%%

\bibitem{Gonzalez:2012mq}
  P.~Gonzalez, S.~Palmer, M.~Wiebusch and K.~Williams,
  %``Heavy MSSM Higgs production at the LHC and decays to WW,ZZ at higher orders,''
  arXiv:1211.3079 [hep-ph].
  %%CITATION = ARXIV:1211.3079;%%

%\bibitem{Heinemeyer:1998yj}
%  S.~Heinemeyer, W.~Hollik and G.~Weiglein,
%  %``FeynHiggs: A Program for the calculation of the masses of the neutral CP even Higgs bosons in the MSSM,''
%  Comput.\ Phys.\ Commun.\  {\bf 124} (2000) 76
%  [hep-ph/9812320].
%  %%CITATION = HEP-PH/9812320;%%

\bibitem{Barate:2003sz}
  R.~Barate {\it et al.}  [ALEPH, DELPHI, L3 and OPAL Collaborations and LEP Working Group for Higgs boson searches],
  %``Search for the standard model Higgs boson at LEP,''
  Phys.\ Lett.\ B {\bf 565} (2003) 61
  [hep-ex/0306033].
  %%CITATION = HEP-EX/0306033;%%

\bibitem{Schael:2006cr}
  S.~Schael {\it et al.}  [ALEPH, DELPHI, L3 and OPAL Collaborations and LEP Working Group for Higgs Boson Searches],
  %``Search for neutral MSSM Higgs bosons at LEP,''
  Eur.\ Phys.\ J.\ C {\bf 47} (2006) 547
  [hep-ex/0602042].
  %%CITATION = HEP-EX/0602042;%%

\bibitem{Carena:2002qg}
  M.~S.~Carena, S.~Heinemeyer, C.~E.~M.~Wagner, G.~Weiglein and ,
  %``Suggestions for benchmark scenarios for MSSM Higgs boson searches at hadron colliders,''
  Eur.\ Phys.\ J.\ C {\bf 26} (2003) 601
  [hep-ph/0202167].
  %%CITATION = HEP-PH/0202167;%%

\bibitem{Ball:2013bra}
  R.~D.~Ball, M.~Bonvini, S.~Forte, S.~Marzani and G.~Ridolfi,
  %``Higgs production in gluon fusion beyond NNLO,''
  arXiv:1303.3590 [hep-ph].
  %%CITATION = ARXIV:1303.3590;%%

\bibitem{ATLAS-CONF-2013-012}
  [ATLAS Collaboration], Note ATLAS-CONF-2013-012.

\bibitem{CMS-13-001}
  [CMS Collaboration], Note CMS PAS HIG-2013-001.

\bibitem{ATLAS-CONF-2013-013}
  [ATLAS Collaboration], Note ATLAS-CONF-2013-013.

\bibitem{CMS-13-002}
  [CMS Collaboration], Note CMS PAS HIG-2013-002.

\bibitem{ATLAS-CONF-2013-030}
  [ATLAS Collaboration], Note ATLAS-CONF-2013-030.

\bibitem{CMS-13-003}
  [CMS Collaboration], Note CMS PAS HIG-2013-003.

\bibitem{CMS-13-004}
  [CMS Collaboration], Note CMS PAS HIG-2013-004.

\bibitem{Beringer:1900zz}
  J.~Beringer {\it et al.}  [Particle Data Group Collaboration],
  %``Review of Particle Physics (RPP),''
  Phys.\ Rev.\ D {\bf 86} (2012) 010001.
  %%CITATION = PHRVA,D86,010001;%%

\bibitem{ATLAS-2012-161}
  [ATLAS Collaboration], Note ATLAS-CONF-2012-161.

\bibitem{CMS-12-044}
  [CMS Collaboration], Note CMS PAS HIG-2012-044.

\bibitem{Aaltonen:2012qt}
  T.~Aaltonen {\it et al.}  [CDF and D0 Collaborations],
  %``Evidence for a particle produced in association with weak bosons and decaying to a bottom-antibottom quark pair in Higgs boson searches at the Tevatron,''
  Phys.\ Rev.\ Lett.\  {\bf 109} (2012) 071804
  [arXiv:1207.6436 [hep-ex]].
  %%CITATION = ARXIV:1207.6436;%%

\bibitem{ATLAS-CONF-2013-014}
  [ATLAS Collaboration], Note ATLAS-CONF-2013-014.

\bibitem{ATLAS-2012-160}
  [ATLAS Collaboration], Note ATLAS-CONF-2012-161.

\bibitem{Schmelling}
  M.~Schmelling,
  Phys.\ Scripta {\bf 51} (1995) 676.

\bibitem{Bechtle:2012jw}
  P.~Bechtle, S.~Heinemeyer, O.~Stal, T.~Stefaniak, G.~Weiglein and L.~Zeune,
  %``MSSM Interpretations of the LHC Discovery: Light or Heavy Higgs?,''
  arXiv:1211.1955 [hep-ph].
  %%CITATION = ARXIV:1211.1955;%%

\bibitem{Aad:2011rv}
  G.~Aad {\it et al.}  [ATLAS Collaboration],
  %``Search for neutral MSSM Higgs bosons decaying to tau^+ tau^- pairs in proton-proton collisions at $\sqrt{s}=7$ TeV with the ATLAS detector,''
  Phys.\ Lett.\ B {\bf 705} (2011) 174
  [arXiv:1107.5003 [hep-ex]].
  %%CITATION = ARXIV:1107.5003;%%

\bibitem{CMS-PAS-HIG-12-050}
  [CMS Collaboration], Note CMS PAS HIG-2012-050.

\bibitem{Behr:2013ji}
  J.~Behr {\it et al.}
  %``Search for Higgs boson production in association with b quarks at CMS in pp collisions,''
  arXiv:1301.4412 [hep-ex].
  %%CITATION = ARXIV:1301.4412;%%

\bibitem{ATLAS-CONF-2012-169}
  [ATLAS Collaboration], Note ATLAS-CONF-2012-169.

\bibitem{CMS-PAS-HIG-12-041}
  [CMS Collaboration], Note CMS PAS HIG-2012-041.

\bibitem{ATLAS:2012txa}
  G.~Aad {\it et al.}  [ATLAS Collaboration],
  %``A search for $t\bar{t}$ resonances in lepton+jets events with highly boosted top quarks collected in $pp$ collisions at $\sqrt{s} = 7$ TeV with the ATLAS detector,''
  JHEP {\bf 1209} (2012) 041
  [arXiv:1207.2409 [hep-ex]].
  %%CITATION = ARXIV:1207.2409;%%

\bibitem{CMS:2012rq}
  S.~Chatrchyan {\it et al.}  [CMS Collaboration],
  %``Search for $Z'$ resonances decaying to $t\bar{t}$ in dilepton+jets final states in $pp$ collisions at $\sqrt{s}=7$ TeV,''
  arXiv:1211.3338 [hep-ex].
  %%CITATION = ARXIV:1211.3338;%%

\bibitem{pythia8}
  T.~Sj\"ostrand, S.~Mrenna and P.~Skands, JHEP {\bf 05} (2006) 026; 
  {\it eadem}, Comput.\ Phys.\ Comm.\ {\bf 178} (2008) 852. 

\bibitem{Ovyn:2009tx}
  S.~Ovyn, X.~Rouby and V.~Lemaitre,
  %``DELPHES, a framework for fast simulation of a generic collider experiment,''
  arXiv:0903.2225 [hep-ph].
  %%CITATION = ARXIV:0903.2225;%%

\bibitem{Cacciari:2008gp}
  M.~Cacciari, G.P.~Salam and G.~Soyez,
  % The anti-k_t jet clustering algorithm
  JHEP {\bf 04} (2008) 063
  [arXiv:0802.1189 [hep-ph]].

\bibitem{fastjet}
  M.~Cacciari, G.P.~Salam and G.~Soyez, Eur.\ Phys.\ J.\ {\bf C72} (2012) 1896 
  [arXiv:1111.6097 [hep-ph]]; 
  {\it eadem}, Phys.\ Lett.\ B {\bf 641} (2006) 57 [hep-ph/0512210].

\bibitem{LHCHiggsCrossSectionWorkingGroup:2011ti}
  S.~Dittmaier, C.~Mariotti, G.~Passarino, and R.~Tanaka (Eds.) [LHC Higgs Cross Section Working Group], 
  %{\sl Handbook of LHC Higgs Cross Sections: 1. Inclusive Observables}, 
  CERN-2011-002, arXiv:1101.0593 [hep-ph].

\bibitem{Dittmaier:2012vm}
  S.~Dittmaier, C.~Mariotti, G.~Passarino, and R.~Tanaka (Eds.) [LHC Higgs Cross Section Working Group],  
  %{\sl Handbook of LHC Higgs Cross Sections: 2. Differential Distributions}, 
  CERN-2012-002, arXiv:1201.3084 [hep-ph].

\bibitem{Arbey:2012fa}
  A.~Arbey, M.~Battaglia and F.~Mahmoudi,
  %``Higgs Production in Neutralino Decays in the MSSM - The LHC and a Future e+e- Collider,''
  arXiv:1212.6865 [hep-ph].
  %%CITATION = ARXIV:1212.6865;%%

\end{thebibliography}
\end{document}